# LIGAND-RECEPTOR INTERACTIONS

**Pierre Bongrand**

Laboratoire d'Immunologie - INSERM U 387, Hôpital de Sainte-Marguerite
BP 29 13274 Marseille Cedex 09 FRANCE
Email : pierre.bongrand@inserm.fr

**Abstract**
          The formation and dissociation of specific noncovalent interactions between a variety of macromolecules play a crucial role in the function of biological systems. During the last few years, three main lines of research led to a dramatic improvement of our understanding of these important phenomena. First, combination of genetic engineering and X ray cristallography made available a simultaneous knowledg of the precise structure and affinity of series or related ligand-receptor systems differing by a few well-defined atoms. Second, improvement of computer power and simulation techniques allowed extended exploration of the interaction of realistic macromolecules. Third, simultaneous development of a variety of techniques based on atomic force microscopy, hydrodynamic flow, biomembrane probes, optical tweezers, magnetic fields or flexible transducers yielded direct experimental information of the behavior of single ligand receptor bonds. At the same time, investigation of well defined cellular models raised the interest of biologists to the kinetic and mechanical properties of  cell membrane receptors.
          The aim of this review is to give a description of these advances that benefitted from a largely multidisciplinar approach.

**Contents**

# 1. INTRODUCTION.

## 1.1 - Aim and scope of the review.

The structure and functions of living cells are critically dependent on the formation and termination of associations between an impressive number of biomolecules. Thus, the cell shape is determined by the organization of a multimolecular scaffold called the cytoskeleton that is made of several tens of protein species whose specific interactions regulate mechanical and topological properties (Pollard, 1994 ; Richelme et al., 1996). The migration of different cell populations through living organisms is dependent on the continuous formation and dissociation of specific bonds between adhesion molecules borne by cells and surrounding tissues. The behavioral response of cells to external stimuli such as adhesive interactions or soluble mediators involves the triggering of a cascade of activation of messenger molecules that will become able to bind to specific receptors scattered through the cells (Bongrand and Malissen, 1998). Thus it is not surprising that Creighton (1993), as quoted by Northrup and Erickson (1992), wrote in his well known treatise on proteins that "the biological functions of proteins almost invariably depend on their direct physical interaction with other molecules".

During the sixties and seventies, a considerable amount of information was obtained on the characterization of many biomolecules with a binding capacity. Many authors reported on the experimental determination of conventional interaction parameters such as affinity constants or kinetic rates of bond formation and dissociation. Much theoretical work was done to achieve correct interpretation of these parameters (Page and Jencks, 1971 ; DeLisi, 1980) and relate them to structural properties of receptors and ligands combined with current knowledge of intermolecular forces (Fersht, 1977 ; Creighton, 1983).

During the following years, at least five major advances gave a new impetus to the study of ligand-receptor interaction :

i) continuous progress in the field of cristallography and biochemistry made available the structure of many ligand-receptor complexes with angström resolution.

ii) Adequate use of site directed mutagenesis allowed to assess the contribution of individual aminoacids to the binding affinity and specificity of protein receptors.

iii) The continuous increase of computer power allowed to take full advantage of the simulation techniques developed during the fifties and develop new procedures. These techniques yielded valuable information on the behaviour of realistic macromolecular systems.

iv) Continous progress in cell biology made it clear that the conventional description of ligand-receptor interaction (through equilibrium and kinetic constants) was insufficient to account for all of phenomena driven by interactions between surface-bound receptors subjected to mechanical stress and imposed displacement.

v) During the last few years, a variety of experimental methods developed by physicists and biologists allowed direct monitoring of ligand-receptor interaction at the single molecule level.

The aim of the present paper is to present an overview of the present situation. Indeed, this opens new research opportunities to physicists that may be willing either to use a physical approach to solve biological problems or to take advantage of biological systems to test physical concepts.

First, we shall briefly provide a general background that may not be familiar to all readers. Then we shall sequentially review recent advances on structural properties of some ligand-receptor couples, new information of the behavior of individual binding molecules, and new theoretical analyses assisted with computer simulations. In each case, we shall present a few examples selected on a quite arbitrary basis rather that aiming at some unattainable completeness. Most examples will refer to proteins, in view of the importance of this class of molecules as well as the author's preference.

## 1.2 - Basic description of molecular associations.

As described in standard textbooks of biochemistry, ligand-receptor interactions might seem a straightforward process liable to fairly simple description. When two molecular species A and B with mutual affinity are mixed in a solution, a time-dependent association between these molecules is expected to occur following the simple equation:

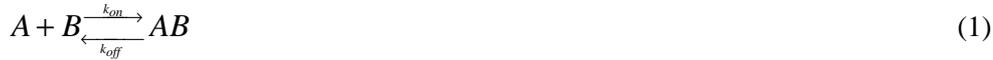

$$A + B \underset{k_{off}}{\overset{k_{on}}{\rightleftarrows}} AB \qquad (1)$$

where the kinetic constants $k_{on}$ and $k_{off}$ account for the forward and reverse reaction according to the following equation:

$$d[AB]/dt = k_{on}[A][B] - k_{off}[AB] \qquad (2)$$

here, the square brackets stand for the concentration of any molecular species, usually expressed in mole/liter. It is readily found by solving equation (2) that, whatever the initial conditions, the system will tend to an equilibrium state following the well-known Guldberg-Waage (or mass action) law. When applied to reactions in solution, this is usually written:

$$[A]_{eq}[B]_{eq}/[AB]_{eq} = k_{off}/k_{on} = K_d = 1/K_a \qquad (3)$$

where "eq" is meant to recall that we are dealing with equilibrium concentrations, $K_a$ is called the affinity constant (in liter/mole) and $K_d$ is called the dissociation constant (in mole/liter).

These simple equations might be considered as a starting point for two main lines of development.

### *1.2.1 - Thermodynamics of binding*.

As pointed out by Williams (1991), "the concept of affinity dominated most thinking about complex biological reactions for many years". Indeed, a major goal consisted of establishing a relationship between the affinity constant and molecular structure of biomolecules. In addition to their conceptual interest, these investigations might be expected to facilitate the design of active drugs or artificial enzymes. In order to fulfil this program, the thermodynamical basis of equation (3) must be discussed (see e.g. Hill, 1960 ; Sommerfeld, 1964a). Provided the concentrations of reagents A, B and AB are low enough, we may write the following relationship:

$$[AB]_{eq}/\{[A]_{eq}[B]_{eq}\} = [AB]°/\{[A]°[B]°\} \exp(-\Delta F°/RT) \qquad (4)$$

here, the superscript ° stands for "standard conditions", this usually corresponds to an *hypothetical* solution of a given species with 1 molar concentration and absence of interaction between these molecules (this amounts to assume that the perfect gas approximation remains valid for 1 molar concentration, see Hill, 1960 ; Gilson et al., 1997).

The quantity $\Delta F°$ is the standard (Helmoltz) free energy of the reaction, this is the variation of free energy caused by combining one mole of A with one mole of B to obtain one mole of complex in an infinite reservoir where A, B and AB are in standard conditions. Finally, R is simply the perfect gas constant (i.e. 8.31 J/°K/mole) and T is the absolute temperature. Since concentrations are equal to 1 mole/litre under standard conditions, equations (3) and (4) may be used to write:

$$K_a = \exp(-\Delta F°/RT) \qquad (5)$$

There is a problem with this expression, since the right hand side is dimensionless. Thus, the correct equation (4) should be used when affinity constants are calculated *ab initio* from basic principles.

Now, $\Delta F°$ may be written as the sum of two contributions (Jencks, 1981):
i) the association between (A) and (B) results in the loss of some degrees of freedom (or only the replacement of free translations and rotations with vibrations). The corresponding contribution to $\Delta F°$ may be denominated as a "connection term" noted $\Delta F°^c$ following Jencks (1981).

ii) The intrinsic contribution of the formation of molecular bonds $\Delta F^i$. Ligand-receptor association involves the formation and dissociation of numerous bonds involving reagents A and B *and solvent molecules*. This may include internal changes of the structure of interacting molecules. More details will be given in the last section of this review.

Note also that in many cases the free enthalpy or Gibbs free energy, G=E+PV-TS, is used instead of Helmoltz free energy, and the enthalpy H=E+PV is used instead of the energy E, when equilibria are studied under constant pressure rather than with constant volume. The product PV of pressure and volume is however quite low in aqueous solution.

A consequence of equation (5) is van't Hoff equation :

$$d\ln(K_a)/dT = \Delta E/RT^2 \qquad (6)$$

(see Weber, 1996, for a discussion of some problems that are often overlooked). Thus, the energy and entropy changes involved in the reaction may be determined by studying the temperature dependence of the affinity constant. Also, the enthalpy may be studied with microcalorimetry.

### *1.2.2 - Kinetics of molecular association*.

Since life works under nonequilibrium conditions, it was warranted to study the kinetics of association between biomolecules. A first point consisted of splitting reaction (1) into the following two steps :

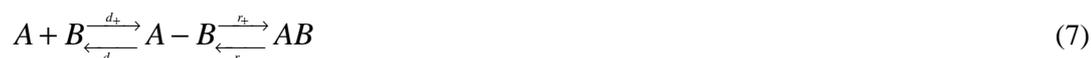
$$A + B \underset{d_-}{\overset{d_+}{\rightleftarrows}} A-B \underset{r_-}{\overset{r_+}{\rightleftarrows}} AB \qquad (7)$$

the first step is the formation of a so-called encounter complex between A and B as a consequence of diffusion. The second step is bond formation. While a theory elaborated by von Smoluchowski (1917) at the beginning of the century is considered as a sound basis for the determination of the rate of molecular encounter, much work was recently devoted to the second step. Further, a simple link between equations (1) and (7) is provided by the widely used (but not-so-easy to prove) steady-state approximation (see e.g. Cantor and Schimmel, 1980 ; deLisi, 1980). Assuming that the concentration of the encounter complex A-B is stationary (i.e. d[A-B]/dt=0), one readily obtains :

$$k_{on} = d_+ r_+/(d_- + r_+) \quad ; \quad k_{off} = d_- r_-/(d_- + r_+) \qquad (8)$$

A notable interest of this concept is that it allowed an extension of the conventional formalism to the domain of surface-attached molecules. This was achieved by George Bell (1978) who elaborated a theoretical framework to account for receptor-mediated cell adhesion. Two major points may be mentioned :

First, in order to study the kinetics of bond formation, Bell separated ligand-receptor association into a diffusion and a reaction phase (equation 7). Further, he suggested that the reaction rate was identical for free and bound molecules, whereas the kinetic constants for the diffusion phase were obtained through a standard Smoluchowski approach, replacing 3-dimensional diffusion with 2-dimensional displacement in the plane of the membrane. Finally, he made use of the steady-state approximation to obtain quantitative estimates for the rate of bond formation between receptor-bearing cells. The limitation of this approach is that it did not account for possible variations of membrane to membrane distance. Thus, it was not suitable to estimate the formation of the first few bonds following cell-to-cell encounter.

Second, a major point emphasized by Bell was that the rate of bond dissociation should be dependent on applied forces. He suggested the following empirical formula :

$$r_-(F) = r_-(0) \exp(\gamma F/kT) = r_- \exp(F/F°) \qquad (9)$$

where F is the applied force, k is Boltzmann's constant, T is the absolute temperature and $\gamma$ is a parameter that should be close to the interaction range of ligand-receptor bonds. Bell estimated $\gamma$ at about 0.5 nm. Although this formula was inspired by experimental data obtained on the rupture of macroscopic material samples (Zurkhov, 1965), it may be somewhat justified with standard theories of reaction rates (see below). Equation (9) proved quite useful since i) it emphasized that bond rupture is a stochastic event, that may occur in absence of distractive force, and ii) it provided an estimate for the force required to substantially enhance the rate of bond formation : using Bell's estimate, $kT/\gamma$ is of order of 10 pN. As will be described below, these concepts were subjected to extensive experimental check during the last few years, and recently some theoretical attempts were done to relate these experiments to results from statistical mechanics.

A thermodynamic approach to the effect of stress on intermolecular association was followed by Dembo and colleagues a few years later (Bell et al., 1984 ; Dembo et al., 1988). Modeling ligands and receptors as Hookean springs (i.e. springs elongating proportionally to the applied force), it is concluded that subjecting a molecular link of length L to a force F will result in a length increase $F/\kappa$ and energy increase $F^2/2\kappa$, yielding an equilibrium constant :

$$K(F) = K(0) \exp(- F^2/2\kappa\, kT) \tag{9}$$

where $\kappa$ is the spring constant. Dembo et al. (1988) further reasoned that there was no thermodynamic necessity implying that bond dissociation rate be increased by a distractive force, and they introduced the concept of *catch-bonds*, whose lifetime would be increase by applied force, in contrast to *slip-bonds*, whose life time should be decreased by disruptive forces, in accordance with intuitive prediction.

Now, in order to provide a quantitative feeling for the parameters we defined, we shall describe several representative examples.

## 1.3 - Typical thermodynamic and kinetic properties of ligand-receptor association.

A prominent example is constituted by antibody molecules that were first obtained by injecting animals with foreign substances called antigens. This procedure induced the synthesis of molecules with a selective capacity to bind antigens used for stimulation. These antibody molecules shared remarkable structural properties allowing them to be included in a family of blood proteins called immunoglobulins. The most abundant immunoglobulins belong to a subtype called immunoglobulin G or IgG. These molecule were observed with electron microscopy by Valentine and Greene (1967) : they appeared as Y-shape assemblies of three rods (about 50 Å length and 40 Å thickness) joined in a fairly flexible region. Each IgG molecule is endowed with two identical antigen-specific binding sites. A typical binding site may be viewed as a cleft of variable depth (5-10 Å), 15-20 Å length and about 10 Å width (Richards et al., 1977), as determined with X Ray cristallography. Antibodies may bind molecules as small as a dinitrophenol group, or large proteins or polysaccharides. The binding sites may involve 5-6 aminoacids or hexose residues (Kabat, 1968).

In a typical series of 21 compiled antigen-antibody couples (Steward, 1977), the affinity constant ranged between $10^4$ and $10^{10}$ M$^{-1}$, although values as high as $10^{12}$-$10^{13}$ were reported by others (Voss, 1993). Association rates displayed relatively restricted variation, ranging between $8\times 10^6$ and $1.8\times 10^8$ M$^{-1}$s$^{-1}$, whereas the dissociation rate varied from $3.4\times 10^{-4}$s$^{-1}$ to 6000s$^{-1}$, thus leading to the common view that antigen-antibody reactions are diffusion-limited, and affinity differences are due to differences in dissociation rates.

In another study, Wurmser et al. (1972) measured the thermodynamic properties of some antibodies specific for protein antigens (albumin or insulin) or carbohydrates (blood group antigens). The affinity constant ranged between $8\times 10^3$ and $6\times 10^8$ M$^{-1}$. The reaction enthalpy and entropy changes ranged respectively between 0 and - 16 kcal/mole and - 35 and 24 cal/mole/°K. Note that the interpretation of older data on antigen-antibody reactions might be somewhat hampered by the heterogeneity of antibody samples. This difficulty was raised by the advent of monoclonal antibody technology.

The range of affinity constants spanned by antibodies is representative of results obtained with other biomolecules. Thus, Lollo et al. (1993) estimated at $10^7$ M$^{-1}$ the affinity of solubilized forms of LFA-1, a cell membrane receptor allowing strong association with ICAM-1, which is another cell surface adhesion molecule. The affinity displayed 100 fold increase upon cell activation : affinity changes related to modification of receptor conformation are indeed a well-known mechanism for the regulation of cell interactions (Pierres et al. 1998a). Lower affinity constants ranging between $10^5$ and $10^6$ $M^{-1}$ were measured on solubilized receptors involved in transient adhesive interactions, such as lymphocyte CD2 (van der Merwe et al., 1993). Conversely, the binding system with highest known affinity is the interaction between avidin or streptavidin (these are proteins of about 60,000 molecular weight) and the small molecule biotin. The affinity constant is of order of $10^{14}$-$10^{15}$ $M^{-1}$ (e.g. Miyamoto & Kollman, 1993). Note that the recent development of surface plasmon resonance based technology proved an incentive to study the equilibrium and kinetic properties of a number of binding systems (Szabo et al., 1995). Also, Sturtevant (1977) reported a compilation of entropy and heat capacity changes associated to a number of ligand-receptor associations, mainly enzyme-substrate binding : $\Delta S$ ranged between -90 and +34 cal/mole/°K.

**1.4 - Inability of the conventional framework to account for biological phenomena.**
The theoretical framework we described in § 1.2.1 is suitable to account for the behaviour of *free* molecules. However, cell function is often regulated by interactions involving *bound* receptors and ligands. In order to illustrate the problems encountered by cell biologists, we shall describe four representative models that recently attracted the interest of many investigators. Then we shall rapidly sketch some theoretical attempts that might be useful to relate cell behaviour to the interaction of individual ligand and receptor molecules.

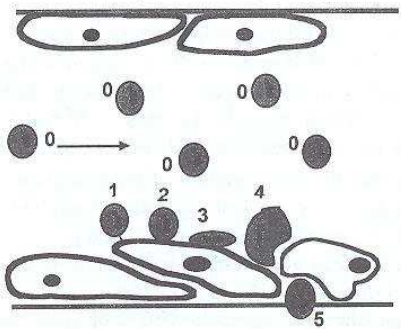

**Figure 1. Adhesive interactions between white blood cells and vessel walls.** When the endothelial cells lining blood vessel walls are activated by an aggressive stimulus such as infection or trauma, they rapidly express selectin receptors. White cells that are flowing with a typical velocity of several hundreds of micrometers per second (0) are then tethered through selectin ligands borne by their membranes (1). Then they begin rolling with 50-100 fold decreased velocity (2) due to the rapid formation and dissociation of selectin-ligand bonds. The following step is a firm adhesion resulting in complete cell arrest (3) due to an interaction between integrin receptors (on white cells) and ligands such as ICAM-1 on endothelial cells. The white cell passage between endothelial cells towards peripheral tissues (4 & 5) also involves the rupture of adhesive interactions between endothelial cells.

*Leukocyte rolling*. The experiments performed *in vivo* by von Andrian et al. (1991) and *in vitro* by Lawrence and Springer (1991) to study the phenomenon of leukocyte rolling certainly played an important role in emphasizing the importance of the lifetime and force dependence of interactions between bound ligands. A major role of white blood cells (i.e. leukocytes) is to patrol throughout living organisms in order to eliminate potentially harmful agents such as pathogens or damaged cells. Thus, if a given tissue is invaded by infectious microorganisms, specific signals will be generated, resulting in the exit of leukocytes from blood towards the site of aggression (Figure 1). Intravital microscopy revealed the basic features of this process, called diapedesis. Leukocytes that are moving with a velocity of several hundreds of micrometers per second first exhibit a spectacular velocity decrease (by a factor of one hundred) and seem to *roll* along the walls of blood vessels (these walls are made of so-called endothelial cells). Second, rolling cells stop completely. In a third step, they exhibit dramatic deformations allowing them to pass through transient gaps appearing between neighbouring endothelial cells, and they reach peripheral tissues.

The molecular basis of this phenomenon was essentially elucidated by Lawrence and Springer (1991) who reconstituted the main features of leukocyte-endothelium interaction in a laminar flow chamber. These authors showed that rolling and arrest were mediated by two separate classes of molecular interactions : leukocytes are endowed with a variety of receptors, including members of the so-called *selectin* and *integrin* families. Rolling is mediated by transient interactions between selectins and their ligands. Bonds can be formed when cells move with high velocity (i.e. several hundreds of micrometers per second, as stated above), they can stand the strong shearing forces generated by blood flow (the wall shear rate is of order of several hundreds of second$^{-1}$). However, they are unable to maintain cells immobile, even if shearing forces are dramatically decreased. In contrast, the interaction between leukocyte integrin receptors and their ligand on endothelial cells may lead to a complete stop. However, integrin/ligand bond do not appear when cells move at physiological velocity in blood flow. A few years later, Patel et al. (1995) made a clever use of genetic engineering techniques to demonstrate that the remarkable property of selectins to form bonds in presence of high shearing forces was abolished when the *length* of these molecules was decreased without altering the binding site. Also, experimental studies performed on isolated blood cell receptors suggest that the affinity constant of both integrins (Lollo et al., 1993) and selectins (Nicholson et al., 1998) is of the order of $10^6$-$10^7$ liter/mole. These experiments strongly suggest that the behaviour of adhesion molecules is not entirely accounted for by their affinity : other properties such as kinetic rates but also binding strength and molecular length may be of importance to regulate their function. This situation could not be ascribed to some complex function of living cells, since rolling could be reproduced either with cells that had been made inert by suitable fixatives (Lawrence and Springer, 1993) or artificial particles that had been coupled with selectins and made to roll along artificial surfaces coated with selectin ligands (Brunk and Hammer, 1997).

*Cell deformability.* Many cell functions are dependent on their capacity to undergo active or passive deformation (Richelme et al., 1996). Much work was devoted to the study of cell mechanical properties. Thus, accurate information on cell viscoelastic properties was obtained by monitoring the deformation of individual cells subjected to controlled aspiration into small micropipettes of a few micrometer diameter (Schmid-Schöenbein et al., 1981 ; Evans and Kukan, 1984). A major issue would be to relate these parameters to the molecular properties of the three-dimensional scaffold called the cytoskeleton. The *in vitro* demonstration by Sato et al. (1987) that the mechanical properties of reconstituted (and simplified) models of the cytoskeleton were dependent on deformation rate suggests that the cell deformability might depend on the lifetime and force dependence of associations between immobilized cytoskeletal components.

*Cell migration.* An important property of living cells is their ability to migrate when they are deposited on suitable surfaces (and possibly stimulated by soluble mediators). As reviewed by Stossel (1993), cell displacement involves the forward emission of a lamellipodium that will adhere to the surface, with subsequent contraction and detachment of the rear part of the cell. Thus, motility is dependent on continuous attachment and detachment. This qualitative concept was made quantitative by Palecek et al. (1997) who studied the migration speed of different cell populations expressing integrin receptors and deposited on surfaces bearing integrin ligands. These authors used different ways of manipulating adhesiveness by varying the surface density of integrin ligands (through standard coupling procedures) as well as cell membrane density and activity of integrins (using genetic engineering techniques). Then, they measured the mechanical strengh of cell-to-surface adhesion by subjecting bound cells to hydrodynamic flow and determining the force required for detachment. Further, they measured cell migration velocity and demontrated a quantitative relationship between the above two parameters. Velocity was maximal for some intermediate value of adhesiveness. This report further emphasizes the physiological importance of the mechanical strength of ligand-receptor bonds.

*Redistribution of adhesion receptors in contact areas.* Recent experimental advances raised the interest of the biological community in the dimension-dependence of affinity constants. It is now well known that most cell membrane molecules may display free lateral diffusion on the cell surface. It is thus not surprising, on a simple thermodynamical basis (Bell et al., 1984), that receptor-mediated adhesion between two cells often results in concentration of binding molecules in the contact area (Kupfer and Singer, 1986). Previous attemps at quantifying this phenomenon (McCloskey and Poo, 1986 ; André et al., 1990) were dramatically improved by Dustin et al.(1996) who deposited cells expressing the CD2 adhesion receptor on supported lipid layers where they incorporated fluorescent derivatives of CD2 ligand (called LFA-3) : Cell to surface encounter resulted in the formation of a contact area where fluorescent molecules were gradually concentrated. This contact area could be visualized with optical techniques such as interference reflexion microscopy (Curtis, 1994). Further work by Dustin (1997) provided a formal proof that concentrated LFA-3 molecules displayed reversible interaction with cell surface receptors, thus supporting the relevance of the concept of two-dimensional binding equilibrium.

The few selected examples we described show that there is a need for quantitative models to account for the relationship between biological phenomena and ligand receptor interactions involving attached molecules. We shall now describe some selected attempts aimed at i) obtaining a workable description of ligand-receptor interaction and ii) using this framework to interpret experimental data.

**1.5 - Models for relating cell adhesive behaviour to molecular properties of their surface receptors.**
Several authors developed quantitative models to relate measurable cell features to the quantitative properties of membrane molecules. Thus, Bell et al. (1984) considered the equilibrium shape of cells bound by specific adhesion receptors. They assumed that the equilibrium contact area ensured minimization of the free energy contributed by i) adhesion molecules diffusing in the plane of cell membranes, with bond formation restricted to the contact area, ii) repulsion between nondiffusible repulsive elements corresponding to bulky macromolecules known to occur on cell surfaces, and iii) stretching of binding molecules to alleviate repulsion. Their model could fit quantitative experimental data on actual adhesion models reported by Capo et al. (1982), but the numerical values of repulsive force and spring constant of cell-cell bridges were fitted parameters. Also, the mechanical properties of the membrane and possibility of active cell deformation were neglected.

A dynamical model was elaborated by Hammer and Lauffenburger (1987) to account for the rate of bond formation between receptor-bearing cells and ligand-coated surfaces. This model proved a suitable framework to account for a number of experimental findings, but it included many unknown parameters such as the contact area, density and accessibility of adhesion receptors or kinetic properties of these receptors. Further models were elaborated to account for specific phenomena such as the aforementioned rolling process (Hammer and Apte, 1992).

Mechanical approaches were elaborated to account for the statics (Evans, 1985 a&b) and kinetics (Dembo et al., 1988) of separation between a cell and a surface. Interestingly, it was demonstrated that adhesion mediated by a few specific bonds could behave as an irreversible process, in accordance with experimental studies (Evans 1985b), and realistic values of fitted parameters might account for some experimental features of the rolling phenomenon (Dembo et al., 1988 ; Atherton and Born, 1972 & 1973). Later experimental studies performed on the detachment of model particles bound to artificial surfaces through anchored adhesion molecules (Kuo and Lauffenburger, 1993) were interpreted within the framework of these models and led to the experimental finding of a linear relationship between the binding strengh and affinity of ligand/receptor bonds.

*In conclusion* the models we briefly described show that there is a need for accurate knowledge of the behavior of anchored ligand and receptors. However, there are too many unknown features to allow an accurate derivation of ligand receptor properties from experimental studies performed on cell-size objects. Further, theoretical knowledge was markedly insufficient to yield accurate prediction of these parameters. We shall now describe three main lines of research that shed a new light on ligand-receptor interaction : i) the development of genetic engineering led to unprecedented accuracy in the understanding of correlations between structural and functional properties of biological receptors. ii) New methodologies allowed direct investigation of ligand-receptor association at the single molecule level, and iii) continuous progress of computer simulation allowed more detailed understanding of the behaviour of complex objects such as protein molecules embedded in aquous electrolyte solution.

**2 - NEW INFORMATION ON STRUCTURE-FUNCTION RELATIONSHIP IN LIGAND-RECEPTOR INTERACTION.**

As previously mentioned, we shall essentially focus on reactions involving proteins, in order to prevent excessive dispersion. The following questions may be considered :

- Is there a preferred kind of interaction (e.g. hydrogen bonds, hydrophobic interaction, salt-links) that is mostly used by proteins to achieve binding affinity and specificity ?
- Is binding affinity contributed by a few strong interactions or many weak bonds scattered on an important area ?
- Does binding require important conformational changes of interacting molecules or may these be considered as rigid ?
- Is there an accurate fit between interacting molecules or are there wide gaps filled with solvent in protein-protein interface ?

In order to address these questions, we shall first describe some properties of representative ligand-receptor complexes that were studied with X ray cristallography. Then we shall review some information obtained by mutagenesis experiments.

**2.1 - General features of protein-ligand complexes**.
The first important parameter may be the contact area. The precise definition of this parameter may not be as straightforward as it might first seem. Great simplification was brought by the concept of solvent accessible surface area (Lee and Richards, 1971 ; Richards and Richmond, 1978). This is the geometrical locus of the center of a spherical probe (considered to represent a solvent molecule, the usual radius is 1.4 Å) remaining in contact with the protein surface. The surface area that is buried in the protein-ligand interface after association provides a convenient measure of the extent of interaction. Recently, Jones and Thornton (1996) studied 59 protein complexes whose cristallographic structure was recorded in the Brookhaven protein database. The reduction of the accessible surface area generated by complex formation ranged between several hundreds and several thousands of squared angstroms. Further, the authors compared the frequency of occurrence of different aminoacid residues in the interface and on the remaining part of the protein surface. A notable conclusion whas that the frequency of nearly all hydrophobic residues was higher in the contact area. Also, they estimated the mean number of hydrogen bonds between reagent surfaces : this was of order of one per 100 Å$^2$. Now, we shall present a few selected examples.

The structure of a complex made between lysozyme and a monoclonal antibody (called D1.3) was studied with 2.8 Å resolution (Amit et al., 1986 ; Mariuzza et al., 1987). The affinity constant was $4.5 \times 10^7$ Mole$^{-1}$. Sixteen aminoacid residues of the lysozyme surface made tight contacts with 17 residues on the antibody combining site. There was a notable complementarity between surfaces, since protrusions occurring on a molecule were matched by depressions on the opposed surface. Twelve hydrogen bonds were identified between surfaces. About 11 % of the lysozyme accessible area (i.e. 748 Å$^2$) was buried during the interaction, together with 690 Å$^2$ on the antibody surface. A later study performed with 1.6 Å resolution led to the conclusion that the cristallographic structure of lysozyme was identical in free and bound molecules. Similarly, no gross difference was found between the conformation the lysozyme-bound monoclonal antibody and free immunoglobulin molecules. The authors concluded that no drastic conformational change was detected in different proteins upon ligand binding. Several years later, the same model was used to compare the conformation of free and bound antibody D1.3 with 1.8 Å resolution (Bhat et al., 1994). This resolution allowed unambiguous localization of water molecules. Twenty three water molecules were bound to the free antibody site, and 48 were localized in the antigen-antibody interface, acting as bridghes between protein surfaces. This was consistent with the experimental finding (obtained with calorimetric studies) that the antigen-antibody association resulted in a net entropy decrease, in contrast with hydrophobic interactions that are supposed to increase entropy as a result of a release of solvent molecules (see § 4.1).

Low affinity interactions (in the ten micromolar range) were also investigated. Garboczi et al. (1996) studied the complex between a particular human T-cell receptor (a specific receptor for foreign structures expressed by a subpopulation of lymphocytes) and its natural antigen, a complex between an oligopeptide of viral origin and major-histocompatibility molecule HLA-A2. The solvent accessible surface area that was buried on the T cell receptor on binding was 1,011 Å$^2$. Twenty hydrogen bonds were identified among a total of 46 interatomic contacts (defined as interatomic distances lower than 4 Å). No gross conformational changes was induced on interacting molecules by complex formation. In another study, Gao et al. (1997) studied the interaction between lymphocyte CD8 molecule and HLA-A2. The total buried accessible area on CD8 was 947 Å$^2$. Interactions were considered as mainly electrostatic since 80 % of atoms were polar in contact regions. Eighteen hydrogen bonds were identified between interacting proteins.The authors suggested that the relatively low affinity value was related to the high number of polar residues in the interface.

Finally, Weber et al. (1989) explored the structural origin of the high affinity ($10^{15}$ Mole$^{-1}$) interaction between the protein avidin and the small biotin molecule. They concluded that binding involved a high number of hydrogen bonds and a conformational change of the protein burying the biotin in a pocked that was closed with a surface loop of avidin.

**2.2. - Information obtained by studying series of mutant molecules**.

As will be detailed below, the difficulty of achieving a quantitative understanding of ligand-receptor association is that the binding free energy represents only a minimal fraction (say a few percent) of the total conformation energy of interacting molecules. Thus, much information could be obtained by comparing series of molecule differing by a few or even a single aminoacid, with might give accurate information on the contribution of a few or even a single molecular interaction to the total binding energy. We give a few selected examples.

The relative importance of electrostatic and hydrophobic interaction was studied by exploring the high affinity interaction between thrombin, a coagulation factor, and hirudin, a 65 aminoacid polypeptide found in medicinal leech. Stone et al. (1989) measured the affinity constant and reaction rate of four mutants obtained by replacing one to four negatively charges glutamic acid residues with glutamine (which amounts to replacing a terminal COO$^-$ group with CONH$_2$). Interaction parameters were measured both in a solution of high ionic force (resulting in efficient screening of electrostatic interactions) and at low electrolyte concentration. The authors tentatively separated the contributions of ionic and non-ionic interactions considered as additive components of the standard free enthalpy of reaction. They found that the non-ionic component $\Delta G°_{nio}$ displayed a similar value of about -15 kcal/mole while the ionic component algebraically increased from -6.9 kcal/mole to -2.1 kcal/mole upon sequential removal of four negative charges. Further cristallographic study demontrated that hirudin bound thrombin at sites both close and distant to the active site (Grutter et al., 1990).

The study of immunological recognition provided many opportunities to demonstrate that the replacement of a few aminoacids in a protein could markedly change binding properties. Thus, while aforementioned monoclonal antibody D1.3 bound hen egg lysozyme with high affinity, no detectable binding was measured on lysozymes from other animal species differing by only 3 or 4 aminoacids (Mariuzza et al., 1987). Further, Chacko et al. (1995) reported a study made on the interaction between lysozyme and a monoclonal antibody (HyHEL-5) : the interface region in the complex contained 23 lysozyme and 28 antibody aminoacid residues. The replacement of a single (positively charged) arginine with a lysine (of similar charge) resulted in the introduction of a water molecule in the interface and concomitant $10^3$-fold reduction of the binding affinity. A similarly exquisite specificity was reported in another model : so-called natural killer cells are endowed with receptors for histocompatibility molecules. These receptors are able to discriminate between histocompatibility molecules from different individuals. The simple exchange of two neighbouring aminoacids (a methionine and a lysine) was sufficient to exchange the specificity of a receptor (Winter and Long, 1997).

Several reports gave some information on the possible functional importance of minimal conformational changes related to complex formation. In a study performed on the lysozyme/antibody model, Hawkins et al. (1993) studied the contribution of residues of the D1.3 antibody to hen egg lysozyme binding. Interestingly, they obtained a mutated molecule with fivefold affinity increase, while none of the altered residues was located in the contact interface. Similarly, Wedemayer et al. (1997) compared the structure of complexes involving antidodies differing by a few residues : A 30,000 fold affinity increased could be achieved by mutations located at distance (more than 15 Å) from the binding site. These mutations seemed to act by stabilizing the conformation displayed by the antibody during binding. When Tulip et al. (1992) studied the cristallographic structures of mutant neuraminidase-antibody complexes, they reported that single sequence changes in some of the neuraminidase residues in the binding site markedly reduced affinity. However, in some cases a sequence change could be accomodated by a structural modification of the conformation of a few residues in the complex (e.g a 2.9 angström shift or a rotation of 150°).

Clackson and Wells (1995) reported a remarkable study on the high affinity ($3\times10^9 M^{-1}$) interaction between human growth hormone (hGH) and a fragment of the hGH receptor bearing the binding site. X ray cristallographic studies revealed that about 30 residues were involved in the interaction on each protein. Controled mutagenesis was used to replace systematically each of these residues with an alanine, which is a relatively small amino-acid ($CH_3CH_2CH(NH_2)COOH$) whose introduction is supposed to remove possible electrostatic or hydrogen bond without adding bulky groups that might reduce affinity through steric repulsion. The authors found that only 9 substitutions (out of 33) on the hGH receptor resulted in marked affinity reduction (between 1 and 4.5 kcal/mole). They suggested that in some cases the desolvation energy and lateral chain reorganization associated to complex formation balanced the energy gain due to bond formation. They also suggested that water molecules might fill gaps between imperfectly matching regions of proteins and form interactions that were fairly isoenergetic with those found in free proteins. They concluded that the dominant importance of a few interactions would facilitate the design of low size synthetic ligand for medical purpose.

Recently, Vallone et al. (1998) used analytical ultracentrifugation to study the association of hemoglobin subunits. They determined the affinity changes generated by simple or double mutation. Also, they took advantage of molecular modeling to determine the accessible surface area variations resulting from these mutations. They concluded that the free energy of burying hydrophobic residues in a protein-protein interface was about -15 cal/mole/$\text{Å}^2$. They also suggested that the contribution of polar interaction to the affinity was low. This conclusion is in line with another report from Davis et al. (1998) who prepared a series of mutant CD2 molecules and studied their low affinity ($\approx$ micromolar) interaction with CD48 ligand. They exhaustively mutated residues located in the CD2-CD48 interface. They concluded that three fairly hydrophobic residus (leucine, phenylalanine and tyrosine) were dominant contributors to the binding energy. Since the affinity constant was independent of the ionic strength, they concluded that the binding free energy was mainly accounted for by hydrophobic interactions, while electrostatic forces might contribute the *specificity* of the interaction.

*In conclusion*, Accurate structural information, is now available at the atomic level on the structure of protein-protein interfaces. Recently, additional data obtained by systematic mutagenesis experiments led to the view that a few interactions might account for most of binding energy. Probably the steric repulsion generated by a single residue or a few unfavorable electrostatic interaction might suffice to prevent binding, thus accounting for the remarkable specificity of many receptors. Available data form a firm basis to test refined theoretical models of protein association. These will be described in the last section of this review.

## 3 - EXPERIMENTAL STUDY OF LIGAND-RECEPTOR INTERACTION AT THE SINGLE MOLECULE LEVEL.

While conventional lines of research on ligand-receptor interaction were followed during the last few years, and indeed methodological advances such as the use of surface plasmon resonance gave a new impetus to the experimental determination of binding rates, a qualitative change in this field was brought by the development of several experimental methods bringing direct information on individual interactions between surface-bound molecules. The interest of this approach is that data interpretation is greatly facilitated when single bonds are monitored, since there is no need to account for the mechanical properties of surfaces and the geometrical arrangement of bonds. Experimental results allowed to check conventional theories with unexpected accuracy. Now, we shall describe these techniques together with selected experimental results. The significance of reported data will be discussed in the final part fo this review. We shall first describe studies made on the determination of bond lifetime and mechanical strength. Then whe shall describe the rather scanty information available on the kinetics of bond formation between surface-attached molecules.

## 3.1 - Lifetime and mechanical strengh of ligand-receptor bonds.

During the last ten years, at least three different methods (Bongrand et al., 1994) became available to monitor the rupture of individual ligand-receptor bonds subjected to distractive forces in the piconewton range. It is now apparent that noncovalent associations between biomolecules can be rapidly ruptured with forces ranging between a few tens and hundreds of piconewtons. Part of the approaches listed below were described in a recent review (Pierres et al., 1998b).

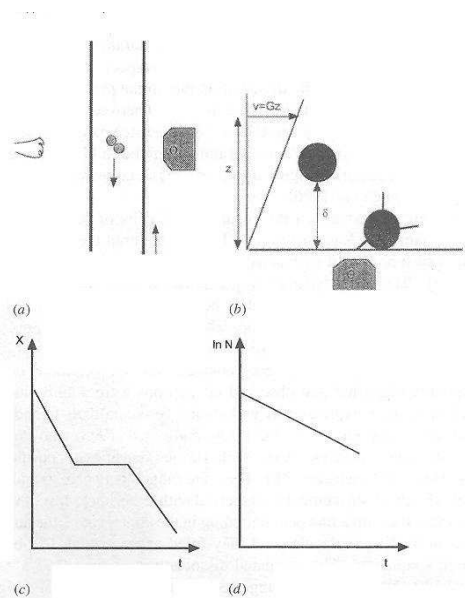

**Figure 2. Studying individual ligand-receptor bonds with hydrodynhamic flow.** The traveling microtube technique developed by H. Goldsmith (A) consists of sending particle suspensions through a vertical capillary tube mounted on a moving stage. Particles are coated with adhesion receptors resulting in doublet formation, and individual doublets are monitored with a microscope (with horizontal axis). The low doublet velocity (typically ~ 25 µm/s) is accurately balanced by the stage motion, thus allowing prolonged observation of the doublet that remains fixed with respect to the microscope field. The hydrodynamic forces exerted on the doublet at the moment of rupture are accurately calculated.

The laminar flow chamber (B) is a parallelepipedic cavity whose floor is monitored with a standard inverted microscope (the objective O is shown). Receptor bearing particles are driven along ligand-coated surfaces with a wall shear rate typically ranging between a few and 100 $s^{-1}$. The force on a bond maintaining a particle arrested may be calculated if the wall shear rate and bond length are known. The distance δ between a flowing sphere and the surface may be derived from the sphere velocity with nanometer accuracy. A typical trajectory is shown when the sphere velocity is plotted versus time (C). Periods of displacement with fairly constant velocity are intersperesed with arrests (horizontal segment) of various duration. The rate of bond dissociation can be obtained by determining the distribution of arrest durations and plotting the number of particles remaining bound at any time t after arrest versus t (D).

**- *Use of hydrodynamic flow*** : a particle of radius $a$ bound to another particle or a macroscopic surface in a fluid of viscosity µ with a locally varying flow of shear rate G is subjected to a distractive force of *order* of $\mu a^2 G$. If bonds are ruptured, particles will then depart with a velocity of order of $aG$ (Figure 2). Thus, if we consider a cell-size particle of 10 µm radius in a fluid of 0.001 Pa.second viscosity such as water, a shear rate of $10 s^{-1}$ may generate an hydrodynamic drag of order of 0.1 piconewton and relative velocity of 100 µm/s. Therefore, the mere observation of the particle with a conventional microscope may in principle allow a detailed examination of single bond rupture with a time resolution of a few tens of milliseconds.

The pioneering studies were performed by Tha et al. (1986) with the so-called traveling microtube apparatus (Figure 2) : they prepared highly spherical red blood cells by exposure to hypoosmotic treatment prior to fixation. These particles were then coated with low amounts of antibodies in order to allow them to bind to each other with a a few or even single bonds. Suspensions were driven through a vertical capillary tube that was mounted on a mobile stage, under continuous monitoring with a microscope whose optical axis was perpendicular to the tube. The stage velocity was adjusted to achieve exact compensation of the displacement of individual particles that could thus be followed for a fairly long period of time. The authors followed the motion of individual doublets that underwent rotation with a sequence of compressive and disruptive hydrodynamic forces. The distribution of force intensities at the moment of rupture could thus be accurately calculated, yielding an estimate of about 24 piconewtons for the binding strengh of the weakest doublets. A comparable estimate of 20 pN was later obtained with an improved methodology based on a cone-and-plate rheoscope allowing rapid variation of the shearing forces (Tees et al., 1993 ; Goldsmith et al., 1994). In later studies, computer simulations were used to extract quantitative estimates from the natural rate of bond dissociation and mechanical strength of bonds. The latter parameter was obtained assuming exponential increase of the bond dissociation rate with respect to force (according to Bell's model as displayed in equation 9, the dissociation rate in presence of a disruptive force F is $r_- \exp[F/F°]$). The dissociation rate of bonds formed between a polysaccharide antigen (blood group B) and specific antibody was 0.04 s$^{-1}$ with a force parameter F° of 35 pN (Tees et al., 1996). Similarly, the interaction between immunoglobulin G and protein G, a natural immunoglobulin ligand of bacterial origin, displayed a dissociation rate of 0.006 s$^{-1}$ and force parameter of 11 pN (Kwong et al., 1996).

Other studies were performed on the separation of cells or particles from surfaces in parallel-plate flow chambers. Kaplanski et al. (1993) monitored the motion of human white blood cells along a surface coated with activated endothelial cells in presence of a very low shear rate (5.25 s$^{-1}$). The hydrodynamic drag exerted on cells interacting with the surface was less that 5 pN. A single bond should thus be sufficient to maintain cells under arrest. Cells indeed displayed numerous transient stops whose duration could be fitted to a theoretical curve obtained with respective values of 0.75 s$^{-1}$ and 0.50 s$^{-1}$ for the rates of bond formation and dissociation in the region of cell-suface contact *after the formation of the first bond*. Inhibition experiments suggested that observed interactions were mainly due to association between E-selectin receptors expressed by endothelial cells and their ligand on white cells. These conclusions were supported by a report from Alon et al. (1995) who studied the motion of white blood cells along surfaces coated with various densities of purified P-selectin, a molecule closely related to E-selectin. The shear rate ranged between 20 and 110 s$^{-1}$. Cells displayed intervals of rapid displacement interspersed with tethering events of various duration. Modeling cells as spheres with transient flattening in contact area, the authors calculated the force exerted on binding molecules and they fitted experimental distributions of arrest durations with Bell's equation. The estimated dissociation rate was 0.95 s$^{-1}$ and the force parameter F° was 120 pN. The authors suggested that the low lifetime and high mechanical resistance of selectin-mediated bonds were a prerequisite to support the rolling phenomenon (see § 1.4 and figure 1). In a later study, Pierres et al. (1996) studied the motion of spherical particles bearing recombinant binding sites of CD48 along CD2-derivatized surfaces. The CD2-CD48 adhesion system is supposed to mediate transient associations between T lymphocytes that scan the surface of different cell populations in order to detect possible abnormalities (Bongrand and Malissen, 1998). The choice of spherical particles allowed the use of a computer-assisted tracking device yielding the coordinates of the centroid of particle images with 20 millisecond and 20 nm accuracy. Also, the force exerted on individual bonds could be accurately calculated. The rates of bond formation and dissociation between surfaces and particles bound by at least one bond were 39.6 s$^{-1}$ and 7.8 s$^{-1}$ respectively. Bell's force parameter F° was 32 pN.

While the flow chamber proved a very convenient tool for studying transient interactions, it was interesting to study the behaviour of molecules supposed to mediate durable bonds. Thus, Pierres et al. (1995) studied the interaction between rabbit immunoglobulin and spherical particle coated with mouse antibodies specific for these rabbit proteins. First, particles were incubated with fluorescent rabbit immunoglobulins, then washed and assayed for fluorescence content after different periods of time. Mean particle fluorescence exhibited minimal decrease after 3h or even 24h incubation, suggesting that binding was durable. Then spheres were driven along ligand-bearing surfaces in the flow chamber with a wall shear rate ranging between 11 and 72 s$^{-1}$. Interestingly, the distribution of arrest durations could be fitted to a two-step interaction model according to the following equation :

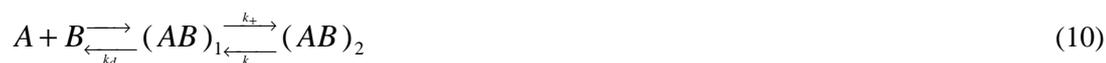

$$A + B \underset{k_d}{\overset{}{\rightleftarrows}} (AB)_1 \underset{k_-}{\overset{k_+}{\rightleftarrows}} (AB)_2 \tag{10}$$

The dissociation rate of the intermediate complex was determined under low dilution conditions (about 3.5 molecules/μm$^2$ on beads and 6,200 molecules/μm$^2$ on the chamber floor). The dissociation rate of the intermediate state $(AB)_1$ was 0.9 s$^{-1}$, with a force parameter of 53 pN. The dissociation rate $k_-$ was of order of 0.01 s$^{-1}$ or less and could not be determined accurately. The stabilization rate $k_+$ was about 0.3 s$^{-1}$ without any significant dependence on the shear rate, as expected. These results show that the concept of a single bound state is only an approximation that may be insufficient to account for the quantitative features of the initial interaction between a moving receptor-coated particle and ligand-bearing surface. It was indeed already known that ligand-receptor association might behave as a multi-step process (Beeson and McConnell, 1994). The above experiment show that this complexity may actually influence particle-to-surface adhesion. In the flow chamber, the relative velocity of the floor and the surface of a sphere is of order of 50 % of the bead velocity, and the velocity of a spherical particle close to the surface is about half the product of the wall shear rate and sphere radius (Goldman et al., 1967 ; see below for more details). Assuming that the length of a ligand-receptor couple is of order of 40 nm, the time available for bond formation between a particle of 1.4 µm radius (Pierres et al., 1995) and the surface in presence of a wall shear rate of 20 $s^{-1}$ is about 6 ms. Thus, provided the detection apparatus is rapid enough, the presented methodology may yield quantitative information on the details of interactions between bound molecules in the millisecond range.

More recently, Pierres et al. (1998c) used the flow chamber technology to study the high affinity interaction between biotin and streptavidin. The lifetime of interactions between biotinylated surfaces and streptavidin-coated spheres was of order of several seconds, i.e. 5-50 fold higher than previously determined on selectin/ligand or CD2/CD48 models. Further, this lifetime was not decreased when the wall shear rate was increased from 10 to 40 $s^{-1}$. However, it was concluded that the flow chamber was not well suited to the study of strong interactions, since i) the chamber floor was rapidly filled with definitively attached particles, which made it difficult to follow a sufficient number of trajectories in a single experiment, and ii) it was somewhat difficult to obtain a reliable distribution of arrest durations, since the computer-assisted apparatus was not adapted to the monitoring of very long arrests.

- *Use of soft vesicles as tranducers* :  this approach was pioneered by Evans et al. (1991). As recalled on Figure 3, the principle consists of approaching with two micropipettes (mounted on micromanipulators) cells or lipid vesicles derivatized with suitable receptor and ligand molecules. After allowing bonds to form, a pipette is pulled out under microscopic control. The applied force results in vesicle deformation, and a spherical shape is recovered when the last bond is ruptured. Thus, the experimental result is the *unbinding force* rather than the bond lifetime. As emphasized by the authors, the interest of this procedure is that the surface tension of the vesicles may be varied in a wide range by controlling the sucking pressure applied through pipettes. Vesicles can indeed be subjected to a distractive force ranging between less than 1 and 100 piconewtons.

The method was used to study the interaction between red blood cells that were cross-linked by a low density of antibodies or lectins (i.e. molecules with an affinity for some cell surface sugars). Since the density of binding molecules was low enough that only a limited fraction of cell-cell encounters resulted in adhesion, it was suggested that attachment was mediated by a few or even a single bond. Surprisingly, the detachment force was of order of 10-20 pN for all tested bridging molecules. The authors suggested that applied forces might uproot membrane receptors rather than rupture ligand-receptor bonds. The vesicle methodology was later improved (Evans et al., 1994 & 1995) by chemically coupling microscopic latex beads to vesicles, and using a piezoelectric transducer to achieve optimal control of pipette position. Finally, an interferometric technique allowed to resolve the bead distance to a flat surface with 5 nm accuracy. Evans et al. (1995) could thus study sphere to surface interactions with piconewton sensitivity. They reported a study of biotin-streptavidin association (Merkel et al., 1995) : when bonds were subjected to a slowly increasing force (100 pN/s), the unbinding force was about 50 pN. This value was 4-5 fold lower than measured with atomic force microscopy (see below), allowing the authors to emphasize the dependence of the unbinding force on the loading rate.

More recently, Chesla et al. (1998) reported fairly accurate determination of bond lifetime with a clever modification of the micropipette technology. The basic idea was to take advantage of a piezoelectric-driven pipette to generate multiple collisions of controlled frequency and duration between immunoglobulin-coated red cells and cells expressing immunoglobulin receptors. This allowed accurate determination of the adhesion probability versus contact duration. The authors emphasized that their methods allowed accurate determination of zero-force association and dissociation rates since the force merely served to provide a signal to the observer. Further, their apparatus certainly allowed piconewton sensitivity. The dissociation rate of bonds formed between human immunoglobulin G receptor and its ligand was 0.37 $s^{-1}$. Since contact duration could be accurately determined, this method might in principle allow to detect possible intermediate binding states.

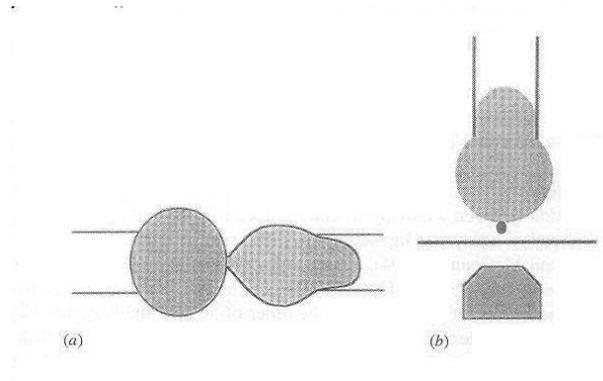

**Figure 3. Studying individual ligand-receptor bonds with biomembrane probes**. The study of adhesive interactions with soft vesicles (red cells or artificial liposomes) was pioneered by E. Evans. A typical experiment (A) consists of using micropipettes to push against each other two vesicles bearing low amounts of receptors and ligands. It may be convenient to use a rigid sphere (e.g. a fixed red cell) and a soft vesicle whose surface tension is accurately controlled by adapting the sucking pressure. When vesicles are progressively separated after contact under microscopic control and video-recording, the deformation of the soft vesicle may be analyzed in order to calculate the force. A refinement of this technique (B) consists of mounting a micropipette on a piezoelectric device (to achieve better control of position) and chemically coupling a small sphere on the vesicle. The distance between this small sphere and a plane surface can be determined with high accuracy by interferometric techniques.

- *Atomic force microscopy* : The principle (Figure 4) consists of moving a ligand-coated surface towards a receptor-bearing tip of a few nanometer thickness mounted on a very soft cantilever (a typical spring constant is about 100 mN/m). The surface is then pulled out, resulting in continuous increase of the distractive force, with continuous monitoring of the cantilever deformation. The rupture of the last bond between surfaces results in a sharp jump of the cantilever, allowing experimental measurement of the so-called *unbinding force*. The cantilever position may be monitored with angström accuracy with optical techniques. The limit set by thermal fluctuations to the force sensitivity is about $(kT/\lambda)^{1/2}$, where $\lambda$ is the spring constant. The reported force sensitivity is of order of 10 pN in liquid medium (Erlandsson and Olsson, 1994 ; Florin et al., 1994 ; Ros et al., 1998). A final point of interest that was noted by several authors (Florin et al., 1994 ; Hinterdorfer et al., 1996) is that interacting molecules do not seem to be altered by the adhesion/rupture cycle, which allows to perform hundreds of cycles on a given position of the microscope tip. Another point is that efficient bond formation may require that the length of adhesion molecules be increased with a chemical spacer (Hinterdorfer et al., 1996) or that one of interacting surfaces be sufficiently deformable (Florin et al., 1994).

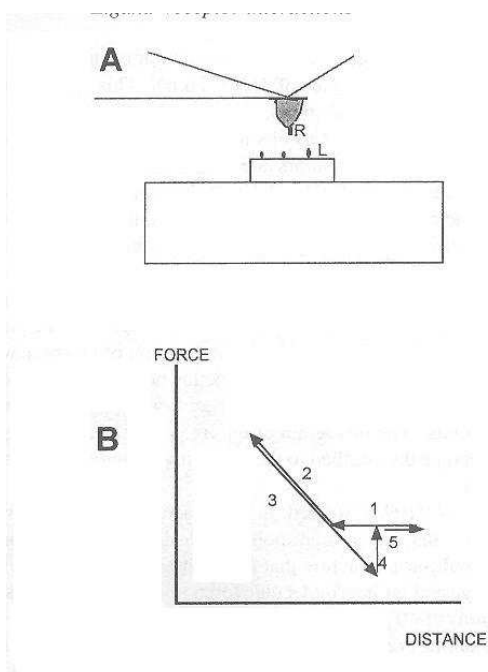

**Figure 4. Studying individual ligand-receptor bonds with an atomic force microscope**. The study of specific biomolecule interactions with atomic force microscopy was pioneered by Moy et al. (1994) and Lee et al. (1994). As shown on Fig. 4A, the piezo-driven surface bearing ligand molecules (L) is subjected to repeated cycles of approach/retraction from the tip derivatized with receptors (R). The tip is mounted on a soft cantilever whose deformation is determined with better than nanometer resolution by optical monitoring. The unbinding force during retraction may be measured by determining the length of the jump (vertical segment 4) occurring during retraction. Actual curves may be more complicated than the very simplified drawing displayed on Fig 4B due to nonspecific forces and occurrence of multiple bonds.

While Hoh et al. (1992) may have detected hydrogen bonds with atomic force microscopy, the application of this apparatus to the study of ligand-receptor interactions was pioneered by Florin et al. (1994) and Lee et al. (1994). The first model studied was the avidin-biotin interaction. A tip was coated with biotinylated albumin and made to interact with soft agarose beads bearing streptavidin binding sites (Florin et al., 1994). Using force scan mode, multiple approach-retract cycles were performed and hundreds of unbinding events could be visualized as sharp jumps of the cantilever (Figure 4). The distribution of unnbinding forces displayed quantized peaks that appeared as multiple of 160 ± 20 pN. This was considered as representative of the detachement force of a single bond. Further, when biotin was replaced with iminotiotin, an analog with 25,000 fold lower affinity to streptavidin, the unit separation force was reduced to 85 ± 15 pN. Interestingly, the authors later determined the affinity constant (i.e. interaction free energy $\Delta G°$), reaction enthalpy $\Delta H°$ (using microcalorimetry) and unbinding force between avidin or the related bacterial streptavidin and biotin or iminobiotin and desthiobiotin analogs (Moy et al., 1994) : they found that the unbinding force was proportional to $\Delta H°$, which led them to define an "effective rupture length" as the ratio between $\Delta H°$ and the unbinding force, yielding a value of 0.9-1 nm. The proportionality between $\Delta H°$ and the unbinding force was confirmed in a later report by Chilkoti et al (1995). The physical significance of interaction parameters will be discussed in the last section of this review.

In other studies, Boland and Ratner (1995) studied the interaction between surfaces coated with adenine and thymine. These basic components of nucleic acids are supposed to bind to each other through two hydrogen bonds. The histogram of frequency of unbinding force suggested the occurrence of quantized peaks ascribed to the separation of individual adenine-thymine pairs. The force was 54 pN.

In another study, Danmer et al. (1995) studied the interaction forces involving a proteoglycan from a marine sponge (this was a large molecule made of a protein core and long polysaccharide chains with a multi-arm structure that is supposed to contribute cell-cell adhesion). Detachment curves suggested an intermolecular force of about 400 pN resulting from of about ten individual interactions of 40 pN each.

In a very interesting study, Nakajima et al. (1997) studied the interaction events between a single molecule of heavy meromyosin (an actin-binding fragment of the actin-binding motor protein myosin) and actin. They estimated at 11.9 milliseconds the half life of a single bond subjected to a disruptive force of 14.8 pN (this was obtained by dividing twice the standard deviation of the unbinding force by the loading rate, i.e. the time derivative of the applied force). Using Bell's formula and taking as a zero-force lifetime a value obtained by Marston (1982) on free molecules (5-100 s), they estimated at 1.7 pN the force parameter $F°$ of the actin-myosin interaction.

In a later study, Vinckier et al. (1998) studied the interaction between Groel, a bacterial chaperone protein (i.e. a protein supposed to stabilize partially unfolded proteins during the early stages of biosynthetic pathways) and several substrates. As expected, interaction forces were greater with unfolded proteins than with native forms. Thus, unbinding forces were 420 pN and 770 pN for native and denatured citrate synthetase enzyme. Further, this force was reduced to 230 pN and 320 pN respectively in presence of ATP that is supposed to modulate GroEL state. Interestingly, the force increased from 440 pN to 620 pN when the cycle frequency was reduced from 1 Hz to 0.1 Hz, while the author reported a decrease of unbinding force when the frequency was higher than 32 Hz. This both emphasized the dependence of unbinding force on loading rate and possible occurrence of weaker intermediate states that might be detected after short contact (i.e. high frequency).

Several studies were devoted to antigen-antibody interactions : Hinterdorfer et al. (1996) reported an unbinding force of 240 pN between human albumin and specific antibodies. Interestingly, they estimated at 1.8 ms the bond lifetime in presence of the disrupting force (this was the ratio between twice the standard deviation 48 pN of unbinding force and the loading rate of 14 nN/s). The natural lifetime of bonds formed between free albumin and antibody was 1500s, i.e. 800,000 fold higher. The force parameter $F°$ was thus 18 pN. Note that the time available for bond formation was estimated at 60 ms. Although this is significantly longer than the contact time in the flow chamber, the measured bond lifetime of 1.8 ms might thus well represent a transient intermediate state, which might hamper the interpretation of F°. In other experiments, Danmer et al. (1996) obtained a rupture force of 115 pN between biotin and anti-biotin antibodies. Allen et al. (1997) found a rupture force ranging between 79 and 1959 pN between ferritin and anti-ferritin antibodies. The force distribution frequency exhibited a period of 49 pN that was ascribed to individual interactions between ferritin and single binding sites of the (multivalent) immunoglobulin molecules. Finally, Ros et al. (1998) compared the unbinding force between fluorescein and the specific sites of two anti-fluorsceyl antibodies differing by a single amino acid (a mutant form resulted from the replacement of an histidin residue with an alanine). The dissociation constants were 0.75 nM and 8.94 nM respectively for the wild-type and mutated sites, and unbinding forces were respectively 50 ± 4 pN and 40 ± 3 pN.

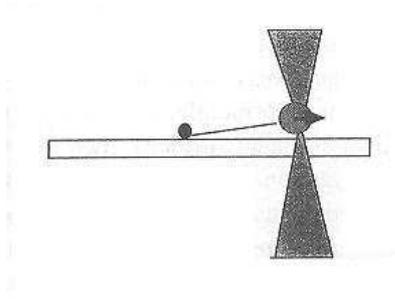

**Figure 5. Sensitive study of individual ligand-receptor bonds with optical tweezers**. The use of optical trapping was pioneered by Ashkin. When a laser beam is focus on the microscope stage, a spherical bead is driven towards the point of focus with a typical force of several piconewtons. In experiments reported by Nishizaka et al. (1995), spheres are bound through long actin filaments to myosin fragments adsorbed on the glass slide : a traction may be exerted on this unimolecular link by moving the microscope stage while the sphere is maintained with the "optical tweezers".

- *Optical tweezers* : the principle of this rapidly developing methodology consists of focusing a laser beam on a small sphere deposited on the stage of a microscope (Figure 5). The force is generated by the deflection of photons composing the diffracted light rays (see Ashkin, 1992) and tends to maintain the bead center on the point where the beam is focused. A notable problem is the potential heating of the bead by the light ray.

This method was used by Nishizaka et al. (1995) who studied the interaction between single actin filaments (1-2 µm length) and heavy meromyosin moieties adsorbed on a glass coverslip. Actin filaments were held by means of microbeads derivatized with gelsolin, an actin binding protein. The force constant of the "laser trap" was estimated at 0.1 pN/nm for a laser power of 95 mW (a clever way of calibrating this force constant consisted of quantifying the brownian motion of beads maintained with a very low power of 0.43 mW, yielding a force constant of 0.44 fN/nm). Actin filaments were subjected to a distractive force under microscopic control. The unbinding force was 9.2 ± 4.4 pN, corresponding to a bond lifetime of about 3 seconds. Using 100-1000 s for the natural lifetime of the interaction (i.e. determined on free molecules ; Marston, 1982), Bell's force parameter F° was estimated at 1.4- 2.1 pN. In a later study, Miyata et al. (1996) reported a mean value of 18 pN for the unbinding force between actin and skeletal muscle alpha-actinin, however, they obtained two force constants when they analysed force/detachment data. Note that this powerful technique was used to determine the force of interaction between receptors of living cells and ligand-coated beads (Choquet et al., 1997).

- *Centrifugation* : in aqueous medium, a typical cell of 5 µm radius and 1,070 kg/m$^3$ density is subjected to a sedimentation force (i.e. weight minus Archimedes force) of 0.4 pN. Thus, a very simple means of probing individual bonds might in principle consist of depositing cell-like particles expressing suitable receptors on ligand-coated surfaces, then inverting culture chambers. These may be subjected to mild centrifugation (say between 1 and 100 g) before assaying detachment. The problem is that ill-defined "nonspecific" interactions often make it difficult to identify the molecular interactions that are being studied. This approach has long been used to quantify cell adhesion (McClay et al., 1991). However, it has only recently been applied to the determination of single bond binding parameters. Indeed, Chu et al. (1994) coated rat leukemia cells with various amount of dinitrophenol-specific antibodies. Cells were then deposited into dinitrophenol-coated culture wells. Plates were reverted after a few minute incubation and count was made of the number of cells remaining bound to chambers for application of a sedimentation acceleration ranging between 1 and 300 g. When the surface density of adhesion molecules was reduced in order that less than 1 % of cells adhered, Application of a sedimentation force of 20-40 pN was required to detach 50 % of adherent cells. Thus, the authors may well have probed individual bonds with this simple approach. More recently, Piper et al. performed an extensive study of the adhesion of colon cancer cells to surfaces coated with aforementioned E-selectin molecules. They measured the surface density of binding molecules on interacting particles, and they extracted the contact area and binding constants of cell-associated receptors as fitted parameters. The estimated dissociation rate varied between 0.35 and 1.8 $s^{-1}$ when the distractive forces increased from 0 to 17 pN. Interestingly, when the force was further increased to 30 pN, the dissociation rate decreased to 0.51 $s^{-1}$, thus raising the intriguing possibility that selectin-ligand might act as *catch bonds* (Dembo et al., 1988) under some experimental conditions.

**- Flexible glass rods as tranducers**. Calibrated thin glass rods provide a simple way of applying small forces to microscopic objects. Kishino et al. (1988) prepared series of rods of increasing flexibility, in order to allow sequential calibration of each rod by the preceding one. They were thus able to calibrate small glass needles with a tip flexibility of 1-10 pN/µm. They coated these beads with myosin in order to make them bind actin. Beads were then manoeuvered with a micromanipulator into contact with isolated actin filaments that had been made fluorescent by exposure to a fluorescent derivative of phalloidin, a high affinity ligand for filamentous actin. Needles were subjected to increasing bending forces under microscopic control, thus allowing quantitative determination of the rupture force of individual actin filaments. The rupture force was 108 pN, and this was increased to 117 pN when actin was incubated with tropomyosin. Also, the motile force generated by individual molecules of the molecular motor myosin was estimated at about 10 pN. A similar approach allowed Meyhöfer et al. (1995) to determine the force required to stop another molecular motor, kinesin, bound to microtubules : this was 5.4±1pN. Cluzel et al. (1996) used flexible optical fibers whose displacements were determined with 10 nm resolution by measuring the deflection of a laser ray. They were thus able to study the extension of single duplex DNA molecules subjected to a force ranging between ~ 10 and 160 pN. Finally, Essevaz-Roulet et al. (1997) determined te force required for mechanical separation of two complementary strands of DNA by monitoring the bending of a glass microneedles. They reported a rupture force of 10-15 pN, and the resolution was sufficient to discriminated between adenine-thymine and stronger guanine-cytosine interactions.

**-Magnetic forces**. A few authors used magnetic beads to subject individual molecules to very small forces. Thus, Smith et al. (1992) studied the extension of single DNA molecules subjected to forces ranging between 0.01 and 10 pN. They were thus able to test the model of freely jointed polymeric chains. A few years later, Strick et al. (1996) studied the elasticity of single supercoiled DNA molecules subjected to forces lower than a few piconewtons.

**- Direct use of thermal fluctuations** : the practical limit of picoforce probing of individual molecules is probably set by thermal fluctuations. However, these fluctuations may also be used to probe molecular interactions. This was clearly done by Liebert and Prieve (1995) who studied the motion of microscopic beads (9 µm diameter) deposited on the stage of a microscope with evanescent wave illumination (Figure 6). It was thus possible to monitor the brownian fluctuations of sphere elevation with nanometer resolution. In addition, these authors subjected beads to controlled vertically oriented radiation pressure. In a typical experiment, 50,000 position measurement were performed at 10 ms intervals, thus yielding an accurate frequency distribution of particle elevation. The interaction potential was easily derived through straighforward use of Boltzmann's law. The authors could thus measure the net weight of the spheres that was estimated at 0.2 pN. In other experiments, they coated beads with immunoglobulin G and surfaces were derivated with protein A, a natural receptor for these immunoglobulin. Their results suggested the occurrence of a specific attractive force with a decay length of 7.8 nm. However, they did not attempt to extract interaction parameters relative to single molecular interactions.

*In conclusion*, there is now a wealth of experimental results on the rupture of single bonds formed between biomolecules. The dissociation rates displayed wide variations, from less than 0.01 $s^-$ $1$ to ~ 10 s$^{-1}$, and forces of several tens of piconewtons were usually required to enhance bond rupture significantly.

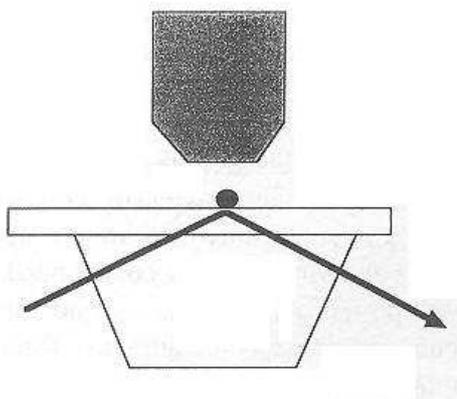

**Figure 6. Sensitive study of particle-surface interaction with evanescent waves**. The use of evanescent waves (sometimes called TIRM for transmission internal reflection microscopy) to determine interfacial forces with hig sensitivity was reported by Liebert and Prieve (1995). A prism is used to send a laser to the glass/medium interface under conditions of total reflection. The region in the upper side of the glass coverslip is thus illuminated with an evanescent wave whose intensity displays exponential decay with respect to the distance z to the interface. Measuring the light scattered by a spherical particle illuminated by an evanescent wave, it was possible to achieve rapid sampling of sphere-to-surface distance with nanometer accuracy. The experimental distribution of sphere-to-surface distance was thus obtained, as a direct illustration of Boltzmann's law.

**3.2 - Direct determination of energy-distance relationship.**

Although the surface forces apparatus does not yield *stricto sensu* information on interactions between individual molecules, this is probably the most powerful tool available for determining the energy-distance relationship between weakly interacting surfaces (Figure 7). Since this was recently applied to the study of specific ligand-receptor interaction, it seemed warranted to describe this methodology. The surface forces apparatus was first developed by Israelachvili and Tabor (1972) for measuring van der Waals forces. The basic principle (Israelachvili, 1992 ; Claesson, 1994) consists of approaching two crossed cylinders coated with silvered (half reflecting) mica surfaces. The distance can thus be measured with angström resolution with an interferometric technique. This apparatus allows to determine the interaction force with about 10 nanonewton accuracy, since a cylinder is mounted on a soft cantilever whose deformation is recorded. When the radius of curvature R of cylinders (usually ~ 1 cm) is much higher than the range of forces between surfaces, it is possible to achieve a direct determination of the relation between distance and interaction energy per unit area W by using the so-called Derjaguin approximation :

$$W = F/2\pi R \qquad (10)$$

were F is the interaction force at any distance between the surfaces. Helm and Israelachvili (1991) were the first to apply this method to the study of specific ligand-receptor interactions by coating mica surfaces with avidin and biotin molecules borne by lipid layers (to impart lateral mobility). They observed a very sharp energy decrease (estimated at 17 kT per molecule) of about 1 angström width. The interaction force was however too strong to allow an accurate study with this technique, due to insufficient cantilever stiffness and rupture of lipid layers. Hence, this method is probably more suited to the study of weaker interactions. Thus, Pincet et al. (1994) could quantify the specific interaction between complementary nucleic bases such as adenine and thymine. Their estimate of about 1.4 kcal/mole for the binding free energy was found consistent with previous estimates. Further, Wong et al. (1997) used this methodology to monitor the association between flexible molecules. A notable finding is the demonstration of a fairly long distance interaction (say several nanometers) between surfaces coated with biological ligands and receptors such as avidin and biotin (Leckband et al., 1992) or antigen and antibodies (Leckband et al., 1995). These could not be ascribed to net electrostatic charges, and the authors suggested that these forces might play a role in steering biomolecules during mutual approach, thus increasing the efficiency of bond formation after a diffusive encounter.

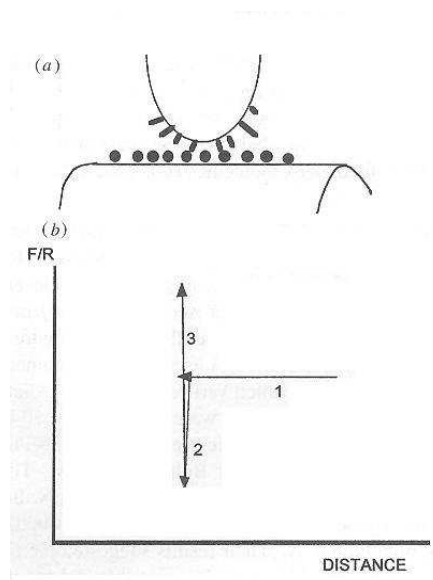

**Figure 7. Study of specific interactions with the surface forces apparatus**. The development of the surfaces forces apparatus and its application to the study of specific interactions (Helm et al., 1991) were pioneered by Israelachvili. The principle (A) consists of approaching two crossed cylinders coated with regular arrays of ligand and receptor molecules. Lateral mobility is required as well as angström smoothness in order to make results interpretable. The remarkable pattern resulting from the occurrence of specific interaction is a sharp jump of the mobile cylinder (when the force is higher than the spring constant) represented as a dotted line (2), and strong repulsion when the distance is further decreased on the angström scale.

**3.3 - Direct determination of the rate of bond formation.**
Few reports were devoted to the experimental determination of the rate of bond formation between surface-attached molecules. Indeed, this is more difficult to study than the induction of bond rupture, since one must in principle induce repeated transient contacts between surfaces, then exert a distractive force and determine whether binding occurred. Also, results are influenced by the length and flexibility of binding molecules as well as their environment, in addition to intrinsinc binding features, which may obscure the significance of experimental data.

Hinterdorfer et al. (1996) studied the interaction between an albumin-coated surface and the tip of an atomic force microscope bearing anti-human serum albumin antibodies. By subjecting the AFM tip to rapid vertical movement and slow lateral displacement, they determined the binding probability (i.e. probability that a binding event with detachment jump occured during a vertical cycle) as a function of position. They concluded that binding could occur when the tip was less than 6 nm distant from an albumin molecule on the surface. They were thus able to estimate the encounter time, yielding a tentative value of the association rate. They estimated the association constant at about $5 \times 10^4$ $M^{-1}s^{-1}$, which was considered as a suitable order of magnitude.

Pierres et al. (1997) monitored the motion of spheres coated with CD48 molecules along planar surfaces derivatized with CD2, a cell membrane receptor for CD48. Particles were driven by a laminar shear flow generating a hydrodynamic force lower than 1 piconewton, i.e. much less than the force usually required to achieve rapid rupture of a single ligand-receptor attachment. Bond formation was thus expected to induce particle arrest. After monitoring hundreds of trajectories, it was possible to determine the frequency distribution of the mean particle velocity on 160 millisecond intervals, including i) all periods of 160 millisecond observation and ii) only periods followed by a binding event. By dividing the number of events in histograms (ii) and (i) for each velocity class, it was thus possible to build a plot of the binding frequency versus average velocity. Then, using a known relationship (Goldman et al., 1967) between particle velocity and distance to the surface, the dependence of binding frequency on "average" sphere-to-surface distance was obtained with about 2nm resolution. Finally, the surface density of binding sites was measured using fluorescent labeling, and the obtained curves were used to extract the binding frequency between a couple of receptor and ligand molecules as a function of the distance between the anchors of these molecules on interacting surfaces : the frequency was inversely proportional to the cube of the distance.

However, as acknowledged by the authors, there were two problems with this approach : First, since the chamber surface was not smooth at the nanometer level, it was difficult to assess the influence of surface asperities on binding frequency. Second, since *average* sphere-to-surface distance were considered, the thermal fluctuations of particle elevation were neglected. This was by no means warranted since the expected amplitude of these fluctuations (i.e. kT divided with net weight) was of order of 100 nm, and substantial fluctuations might occur within a 160 ms interval.

These difficulties were addressed in a later study (Pierres et al., 1998d) made on the homotypic interaction between spheres and surfaces coated with binding sites of cadherins, which are important cell adhesion receptors (homotypic means that the receptor is its own ligand). The chamber floor was an atomically smooth mica sheet where recombinant cadhering moieties bearing terminal hexahistidine tags were bound through nickel ions. Thus, the surface was expected to be smooth at the nanometer level. The authors performed numerical simulations of vertical brownian fluctuations to derive the distribution of sphere-to-surface distance from mean velocity during 160-millisecond intervals. This required an experimental determination of sphere-to-surface interaction : this could be achieved by analyzing the particle sedimentation rate, which allowed quantitative force determination with a sensitivity of order of 10 femtonewtons. Using an experimental determination of the surface density of binding sites on interacting surfaces, the estimated that the rate of association between a pair of attached adhesion receptors was of order of $1.2 \times 10^{-3}$ $s^{-1}$ with an interaction range of order of 10 nm. Note that the significance of the interaction range is difficult to assess, since it is not obvious to determine how the hydrodynamic radius of macromolecule-coated beads might be evaluated. Noticing that the effective concentration of a molecule in a sphere of 10 nm radius is $4 \times 10^{-4}$M, the corresponding association rate is 3 $M^{-1}s^{-1}$.

The problem with both aforementioned approach is that it is difficult to know whether the authors determined an association rate or the frequency of encounters between attached receptor and ligand molecules with a configuration consistent with association.

Recently, Chesla et al. (1998) reported a clever extension of micropipette-based methods allowing fairly direct determination of forward and reverse kinetic rate constants. This consisted of inducing repeated transient contacts between receptor- and ligand- bearing cells, and calculating the probability that an encounter might result in bond formaction, thus yielding resistance to a weak distractive force. This was found to probe predominantly single ligand-receptor bonds. This allowed fairly direct determination of the product between a two-dimensional forward rate constant and contact area. The authors studied the interaction between surfaces coated with human immunoglobulin G and specific receptors respectively. Following quantitative determination of the surface densities of binding sites, they estimated at $2.6 \times 10^{-7}$ $\mu m^4 s^{-1}$ the product between the two-dimensional association rate and contact area. Estimating this contact area at about 0.1 $\mu m^2$, they could thus obtain a 2-dimensional rate of bond formation of $2.6 \times 10^{-6}$ $\mu m^2 molecule^{-1} s^{-1}$.

## 4 - THEORETICAL ANALYSIS OF LIGAND-RECEPTOR INTERACTION.

We have now reviewed quantitative information on i) experimental features of interactions between free or surface-bound molecules and ii) structural properties of regions of contact between bound biomolecules. In order to relate these data, we need first, to find a link between intermolecular forces and structural properties of a given ligand-receptor couple, second to relate intermolecular forces to measurable association or dissociations contants. These steps will now be followed sequentially.

### 4.1 - Intermolecular forces.
A detailed description of intermolecular forces would not fall into the scope of this review, and we refer the interested reader to the numerous remarkable treatises that have been devoted to this topic (Hirschefelder et al., 1954 ; Margenau and Kestner, 1969 ; Maitland et al., 1981 ; Israelachvili, 1991). Thus, we shall only recall very basic information before giving a brief sketch of present day situation.

***4. 1. 1 - Conventional description of intermolecular forces***. Atoms or molecules are clearly subjected to electromagnetic interaction, and quantum chemistry provides a very powerful formalism to deal with these phenomena. Indeed, as written by Eyring et al. (1944) "in so far as quantum mechanics is correct, chemical questions are problems in applied mathematics". However, there is certainly a need for a simplified language to help convey an intuitive feeling for intermolecular forces. We shall hus consider four types of interactions that are often treated as distinct entities, although they are not altogether independent.

   ***Electrostatic interaction between charged molecules***. In vacuum, the interaction energy between two electric charges q and q' separated by a distance r is :

$$V = qq'/4\pi\varepsilon_0 r \tag{11}$$

if charges q and q' are expressed in units of an electronic charge ($1.6 \times 10^{-19}$ C), r is in angström, and V in kcal/mole, equation (11) yields :

$$V = 331 \, (qq'/r) \tag{12}$$

*Influence of solvent.* In a material medium, the electric field generated by a distribution of fixed charges with volume density $\rho$ will induce a polarization of the surrounding molecules, leading to a volumic density of dipole moments **p**. The basic assumption is that **p** is proportional to the electric field **E** following :

$$\mathbf{p} = (\varepsilon - \varepsilon_0) \, \mathbf{E} \tag{13}$$

were $\varepsilon$ is the dielectric constant of the medium (note that we use SI units). Now, since the electric field generated by a dipole **p** at any point of space at distance r is $(1/4\pi\varepsilon_0)$ **grad**$(-1/r)$, it is easily shown that the electric field generated by the dipole field is equivalent to the field generated by a volume distribution of charges - div **p** (combined with a surface distribution of density **p.n**, where **n** is the unit vector normal to the dielectric surface, if this does not fill the entire space). Thus, the standard Poisson equation is replaced with :

$$\text{div } \mathbf{E} = (\rho - \text{div } \mathbf{p})/\varepsilon_0 \tag{14}$$

assuming the dielectric constant is uniform and combining (13) and (14), we obtain :

$$\text{div } \mathbf{E} = \rho/\varepsilon \tag{15}$$

The field and potential are thus reduced by a factor $\varepsilon_r = \varepsilon/\varepsilon_0$ defined as the relative dielectric constant of the medium. The remarkable, and well-known, property of water is that the relative dielectric constant is very high, about 78 at room temperature, thus resulting in drastic reduction of ionic interactions. However, it is important to know to what extent this macroscopic concept is relevant to short-range molecular interactions. It is thus necessary to discuss the relationship between the macroscopic dielectric constant and the microscopic properties of a material medium (see Bongrand, 1988 for more details). When a molecule is exposed to a *sufficiently weak* electrostatic field **E**, it acquires an average dipole moment **P** proportional to **E** according to the following formula :

$$\mathbf{P} = \alpha \, \mathbf{E} \tag{16}$$

where $\alpha$ is the polarizability. The polarization of a material medium in presence of a macroscopic field **E** is readily calculated by noticing that the effective field experienced by a spherical molecule is equal to the difference between **E** and the contribution of the molecule to the average field, yielding **E** + **p**/3 $\varepsilon_0$. It is thus possible to relate the macroscopic dielectric constant to the volume density N and polarizability of individual molecules, leading to Clausius-Mosotti formula (Sommerfeld, 1964b) :

$$(\varepsilon-1)/(\varepsilon+2) = N\alpha/3\varepsilon_0 \qquad (17)$$

Now, the polarizability is the sum of two terms :
- $\alpha_i$ is an induction term due to the polarization of electronic clouds by external fields.
- $\alpha_p$ is due to the orientation of permanent dipole moments in an external field. Assuming that pE/kT is much smaller than 1, where p is the permanent dipole moment, k is Boltzmann's constant and T is the absolute temperature, we obtain :

$$\alpha_p = p^2/3kT \qquad (18)$$

If we consider liquid water, the contribution of the induction term to the dielectric constant is low. Indeed, the induction polarizability of water is less than that of methane ($CH_4$, a molecule of comparable size), and the relative dielectric constant of alcanes is of order of 2 (Weast, 1986). Further, the permanent dipole moment of the water molecule is lower than that of NOCl, while the dielectric constant of water is fourfold higher. The origin of the high dielectric constant of water is indeed the high correlation between the orientations of neighbouring water molecules (Eisenberg and Kauzmann, 1969). Thus, there are three reasons for questioning the use of macroscpic dielectric constant to estimate intermolecular forces :
i) near a charged species, the electric field may be higher than kT/p, leading to a reduction of the effective polarizability (this is dielectric saturation).
ii) Near a protein-protein interface, the solvent structure may be altered with concomitant modification of the correlation between the orientation of neighbouring molecules.
iii) If space is not filled with water, even if a macroscopic dielectric constant can be used, the electric field can only be determined by solving field equations with proper account for the geometry of the solvent-accessible region.

This explains the introduction by different authors of "effective dielectric constants" displaying a wide variation range, between about 2 and 78. Also, the dielectric "constant" was sometimes replaced with a function of distance (see. e.g. Warshel and Åqvist, 1991).

*Influence of surrounding ions.* Dissolved electrolytes may efficiently screen electrostatic interactions in aqueous solution. The standard procedure consists of combining Boltmann's equation with standard electrostatic treatment, leading to the so-called Poisson-Boltzmann equation :

$$- \text{div } \mathbf{grad}\, V = \left[\rho + \sum_i q_i c_i \exp(-q_i V/kT)\right]/\varepsilon \qquad (19)$$

where the summation is extended to all charged species of concentration $c_i$ and charge $q_i$. If the potential V is much smaller than kT, equation (19) yields the so-called linearized Poisson-Boltzmann equation. Now, if the dielectric constant is nonuniform, we obtain, by combining equations (13-15) :

$$- \text{div } (\varepsilon\, \mathbf{grad} V) = \rho + \sum_i c_i q_i \exp(-q_i V/kT) \qquad (20)$$

As recently reviewed (Sharp and Honig, 1990 ; Honig and Nicholls, 1995), the increase of computer power led to a renewal of interest in this classical electrostatic approach, since it became feasible to calculate the potential field around a realistic molecule with the finite difference method. The surface may be defined as the solvent-accessible surface and determined on the basis of X ray crystallography. The molecular charge distribution may be obtained with refined quantum mechanical calculations. The relative dielectric constant is of order of 2-4 within proteins and 78 in the solvent accessible area. The validity of these calculations may be checked, e.g. by considering the effect of local potential on the dissociation constant of ionizable groups (Honig and Nicholls, 1995).

As a consequence, present day calculations are performed by combining numerical resolution of Poisson-Boltzmann equation and simulation of some individual solvent molecules. More details will be given below.

***- Hydrogen bonds.*** The importance of the hydrogen bond was emphasized by Pauling (Pauling, 1960). A major example is the water molecule : due to their electronegativity, oxygen atoms bear a net negative charge and hydrogens a positive charge. It is therefore not surprising that an attractive force might exist between an hydrogen atom of a water molecule and the oxygen of a neighbouring molecule. The high hydrogen bonding capacity of the water molecule is responsible for the high boiling temperature of water (i.e. 100°C) as compared to similar molecules such as $H_2S$ (-60°C). Some authors argued that the hydrogen bond could not be considered as purely electrostatic, but some partial covalent bond might be responsible for a strong dependence on orientation (see e.g. Schuster, 1978). The energy of a typical hydrogen bond is of order of 5 kcal/mole. It is likely that hydrogen bonding is responsible for some aspects of the behaviour of water : this is an highly structured liquid, with a contribution of tetrahedral ice-like configurations where a molecule is bound to four neighbours with hydrogen bonds. The problem with ligand-receptor interactions is that the formation of an hydrogen bond between two solute molecules is likely to result in the release of water molecules bound to hydrogen donor and acceptor groups. This point was very elegantly demonstrated in a recent report by Connelly et al. (1994). These authors studied the interaction between immunosuppressive drugs tacrolimus or rapamycin and the cell protein receptor FKBP-12. They studied the structure of the complex with X-ray cristallography, achieving 1.4 Å resolution. Using site-directed mutagenesis, they removed an hydroxyl group on a tyrosine (that was thus replaced with a phenylalanine). They used binding and microcalorimetric studies to compare the behaviour of wild-type and mutated proteins. Their findings illustrate the complexity of molecular interactions in aqueous environment. Mutation altered the interaction free energy of FKBP-12 with rapamycin by only 0.8 kcal/mole, which reflected a complex balance : two hydrogen-bound water molecules of the unliganded wild-type protein were released on binding, resulting in enthalpy and entropy increase. Since these water molecules were absent on the mutated protein, the reaction enthalpy was lower by 3 kcal/mole on the mutated protein, but the entropy was also lower, resulting in the modest affinity loss of 0.8 kcal/mole.

***-The $r^{-6}$ attraction and short-range repulsion.*** Three kinds of interactions may be responsible for an intermolecular attractive energy inversely proportional to the sixth power of separation distance :

- The average interaction between two freely rotating dipoles separated by a distance r is readily calculated by weighting orientations with Boltzmann's factor, which yields :

$$W_K = - (p_1^2 p_2^2 / 24\pi^2 \varepsilon^2 kT)/r^6 \qquad (21)$$

where $p_1$ and $p_2$ are the dipole moments of interacting molecules. For historical reasons, this is called Keesom interaction. For two water molecules, if r is expressed in Angström, the numerator of (21) is equal to 2600 kcal/mole in vacuum. When dipolar molecules interact in aqueous medium, an "effective dielectric constant" must be used. This depends on the frequency of molecular rotation.

- The interaction energy between an electric field E and a molecule with polarizability $\alpha$ but no permanent dipole moment is equal to $-\alpha E^2/2$. If the electric field is generated by a freely rotating molecule with dipolar moment p, the average energy obtained by integrating over all spatial orientations without any Boltzmann factor (this is the high temperature limit) is :

$$W_D = - \alpha p^2 / 4\pi\varepsilon^2 r^6 \qquad (22)$$

This is called Debye interaction. The numerical coefficient for two water molecules *in vacuum* is 138 kcal/mole when r is in angstrom. This value includes the contributions of both dipole moments to the interaction.

- It is intuitively reasonable to consider that even apolar molecules may express rapidly fluctuating dipole moments resulting from the displacement of individual electrons, which might result in attractive forces. However, a quantum mechanical treatment was required to obtain a quantitative account of this phenomenon. This was first performed by London (1930) who used second order perturbation theory to estimate the so-called dispersion energy of interaction between two molecules (1) and (2) at distance r :

$$W_L = -(3/2)[E_{I1}E_{I2}/(E_{I1}+E_{I2})]\, \alpha_1\alpha_2\, /16\, \pi^2\varepsilon^2 r^6 \qquad (23)$$

where $E_{Ii}$ stands for a characteristic transition energy that is often approximated as the first ionization energy. Using 12.6 eV for this constant, the numerical coefficient of $r^{-6}$ for the interaction between two water molecules is 445 kcal/mole in vacuum, when r is expressed in angström. Note that this coefficient is independent of temperature.

Thus, there is some theoretical justification for the occurrence of an interaction energy $-A_{ij}/r^6$ between two molecules (i) and (j) separated by a distance r. This overall attraction is often called van der Waals interaction  Further, equations 21-23 favour the following approximate combining rule :

$$A_{ij} = (A_i\, A_j)^{1/2} \qquad (24)$$

*Short distance repulsion and Lennard-Jones potential.* It has long been observed that interatomic approach is ultimately hampered by rapidly increasing repulsive forces (once called Born repulsive forces, see e.g. Böhm and Ahlrichs, 1982, for a quantum mechanical treatment). Since many aspects of molecular behaviour are not highly dependent on the details of the force/distance relationship, it was often found convenient to account for $r^{-6}$ attraction and short-distance repulsion with the empirical 6-12 or Lennard-Jones potential :

$$W_{LJ} = -4\varepsilon\, [(\sigma/r)^6 - (\sigma/r)^{12}] \qquad (25)$$

The main feature of this expression is the occurrence of a sharp minimum for the energy (W=-$\varepsilon$ for r= 1.12 $\sigma$) followed by rapid energy increase when the distance is decreased (W is zero for r=$\sigma$ and ≈ 10 $\varepsilon$ when r=0.8$\sigma$) and the interaction rapidly becomes negligible at high distance (W=-0.08 $\varepsilon$ at r=1.5$\sigma$). As a consequence, two unbonded groups coming into close contact will tend to remain separated by a fairly fixed distance depending only on their structural properties.

This sharp energy/distance variation gave some support to a highly simplified view of short distance intermolecular forces (including hydrogen bonds, $r^{-6}$ attraction and "Born" repulsion) : atoms may be viewed as rigid spheres (whose radius is called the "van der Waals" radius) with a constant interaction energy. Experimental values of van der Waals radii have long been obtained by determining the minimal distance between a given pair of atoms in protein cristals, when they are not linked by a covalent bond (see. e.g. Creighton, 1993).

The simplest predictive scheme for interaction energies in liquid media may well be a model reported by Eisenberg and McLachlan (1986) : The predicted solvation energy involved in the transfer of a given residue from the protein interior to aqueous environment is simply the product of the solvent accessible area (as defined by Lee and Richards, 1971) and a characteristic free energy that was respectively estimated (in cal/mole/Å$^2$) at 16, -6, -24, -50 and 21 for carbon, neutral O or N, O$^-$, N$^+$ and sulphur. A similar scheme was devised by Ooi et al. (1987) who used experimental values of thermodynamic parameters of transfer of 22 model compounds from an organic liquid to water. They were thus able to obtain solvation parameters for 7 typical groups (e.g. aliphatic carbon or hydroxyl), and they used these parameters to estimate the free enthalpy of solvation of 21 compounds that had not been used for parameter fit : the root mean square deviation between theoretical and experimental values of ΔG was about 2 kcal/mole.

*In conclusion*, the conventional approach we describe led to a fairly simple view of ligand-receptor binding energies as a sum of long-distance electrostatic interactions, liable to conventional electrostatic description, and highly complex short-distance forces, that may be represented as contact interactions between fairly rigid atoms. This view emphasizes the importance of the geometric particularities of interacting molecules, and makes understandable the potential influence of molecular flexibility.

However, this simplified view is not entirely satisfactory. Indeed, ligand-receptor binding energy usually represent only a few percent of total conformation energies. Thus, highly accurate methods are required to provide a quantitative account of thermodynamic binding parameters. We shall now describe present day approach, whose notable feature is a more and more extensive use of computer simulation. Then we shall add a few remarks on the widely used concept of "hydrophobic interaction".

*4.1.2 - Accurate quantitative study of interaction energies*. The most powerful strategy presently available for studying molecular interactions consists of using high speed computers to simulate the behaviour of a few protein molecules in aqueous environment. This approach conceptually requires two steps :
- First, quantitative expressions for molecular forces must be derived. Accurate determination of the interaction between small molecules (such as water) have been achieved with quantum chemical approach for more than ten years (see e.g. Szezesniak and Scheiner, 1984, and Böhm et al., 1984, for typical examples, and Szabo and Ostlund, 1982, for a textbook presentation of quantum chemistry). However, these complex calculation schemes cannot be applied to realistic systems for lack of computer power, and sets of simplified energy/distance functions are in current use. An important point (Warshel and Åquist, 1991) is that these potential functions must be calibrated with the experimental values of the parameters they are supposed to reproduce (e.g. solvatation energy) rather than directly derived from quantum mechanical calculations.

A typical example is the OPLS (optimized potentials for liquid simulation ; Jorgensen and Tirado-Rives, 1988) set of functions. Parameters were reported for 25 peptide residues and common neutral and charged terminal groups. The interaction energy between unbound groups was represented as the sum of an electrostatic term and a Lennard-Jones potential. Each individual atom was considered as a an interaction site, except $CH_n$ groups that were treated as united atoms. Water molecules were described with standard TIP4P model (involving similar van der Waals and electrostatic terms ; Jorgensen et al., 1983). Two notable points are that i) no special term was found necessary to account for hydrogen bonds and ii) the relative dielectric constant was taken equal to 1.

A second step consists of feeding a standard computer program such as AMBER or CHARMM (Brooks et al., 1983) with protein structures and potential functions. The dynamic evolution of the system can thus be simulated by a series of steps. Technical details may be found in the paper by Brooks et al., 1983) and the review by Beveridge and DiCapua (1989) for the problem of free energy determination.

This approach was used by Miyamoto and Kollman (1993) to compare the free energy of interaction between streptavidin and biotin or two analogs, thiobiotine and iminobiotine (obtained by replacing an oxygen atom by a sulphur or NH group respectively). The simulated system was a box containing biotin and streptavidin surrounded with 502 water molecules. A 2 femtosecond step was used. The calculated relative binding free energies were 3.8 and 7.2 kcal/mole, thus comparing well to experimental values of 3.6 and 6.2 kcal/mole for thiobiotin and iminobiotin. A semiquantitative agreement was obtained for the binding free energy of biotin (-22 to -24 kcal/mole calculated range, with an experimental value of -18.3 kcal/mole). However, the problems raised by the determination of association free energy (including entropy penalty) in order to relate the obtained free energy to the standard binding free energy will be discussed below.

*4.1.3 - Some remarks on the so-called hydrophobic bond*. Although the denomination of "hydrophobic bond" might be somewhat criticized (Isrealachvili, 1991), there is some justification in the use of this term to describe the attraction observed in aqueous medium between molecules composing macroscopically hydrophobic substances, i.e. liquids that are non miscible with water or solids on which a deposited water droplet does not spread. The basic idea is that the high hydrogen bonding ability of H$_2$0 is responsible for the remarkably high surface tension of liquid water (about 72 mJ/m$^2$) as compared to apolar substances such as liquid alcanes (about 20 mJ/m$^2$). Remarkably, the interfacial free energy of e.g. water/hexane (51 mJ/m$^2$) is comparable to the free energy *per unit of accessible area* involved int the transfer of protein components from the molecule interior to the solvent (Richards and Richmond, 1978 ; this energy was estimated between 20 and 33 cal/mol/Å$^2$, i.e. 14 and 23 mJ/m$^2$). However, detailed thermodynamic analysis of this interaction shows that it is fairly complex and dependent on the structure of molecules interacting with water. This point was emphasized by Lee et al. (1984) who studied the structure of water near *extended* planar hydrophobic surfaces : many water molecules kept an hydrogen oriented towards the surface, and the density profile of water exhibited marked oscillatory behavior. This is in contrast with the structure of water near a *small* apolar structure : in this cases, solvent molecules formed rigid chlatrate-like (i.e. cage-like) structures surrounding the apolar structure, with high rigidity and high mutual association, resulting in high free energy, with negative enthalpy and negative entropy changes.

The force/distance law was also investigated : both theoretical approaches (Pratt and Chandler, 1977) and computer simulation (Pangali et al., 1979) suggested that the interaction energy between small apolar molecules might exhibit oscillary behavior, with a first minimum corresponding to molecular contact and a second miminum corresponding to the intercalation of a water molecule between apolar structures (Zichi and Rossky, 1985). More recently, Martorana et al. (1997) performed a simulation of the hydrophobic interaction between a few (5 or 6) Lennard-Jones spheres in a box containing 722 or 723 water molecules represented with TIP4P potential. The attraction was oscillatory with two energy minima at 3.5 and 4.5 Å separation. The maximum attraction might reach about 70 pN. Interestingly, interactions between spheres were not additive, and the range of forces was markedly enhanced when multibody interactions occurred. This is in line with an experimental report by Israelachvili and Pashley (1982) : using a surfaces forces apparatus, they were able to detect a long range attraction between hydrophobic surfaces, with an attraction energy of 22×exp(-D/10) mJ/m$^2$, where the distance D is expressed in angström. Interestingly, the interaction range was markedly decreased by the presence of hydrophilic patches. These results were recently confirmed with atomic force microscopy (Tsao et al., 1993).

The lack of additivity of intermolecular forces (Margenau and Kestner, 1969) must be kept in mind, since this may make meaningless any attempt at obtaining an intuitive understanding of intermolecular forces in term of additive simple components.

*In conclusion*, it is now possible to estimate with high accuracy the thermodynamic changes resulting from molecular contact between receptor and ligand molecules in aqueous environment. In the last part of this review, we shall describe the models that are available to provide a link between these molecular data and macroscopic rates of bond formation and dissociation.

**4.2 - Link between intermolecular forces and thermodynamic or kinetic reaction parameters.**

We shall sequentially discuss association rates, dissociation rates and equilibrium constants.

*4.2.1 - Association rate.*

It was often found convenient to split the association between molecules A and B into a diffusion and a reaction step, as previously described. This discrimination is useful in order to compare the behaviour of free and attached molecules. As was made clear in a very interesting paper by Shoup and Szabo (1982), there is some arbitrariness in the separation of these steps. This point will be dicussed below.

The standard diffusion problem may be stated as follows : considering a solution of free molecules A and B with diffusion constants $D_A$ and $D_B$, we wish to calculate the rate at which two molecules A and B will approach within a "binding distance" R of order of the sum of the linear sizes of these molecules.

*A naive simplified view*. We consider a molecule A moving with velocity *v* in an assembly of randomly distributed fixed molecules B. If the radius of curvature of the trajectory is much higher than R, the number of collisions during time *t* will be ~ $\pi R^2 vt$ [B], where [B] is the number of B molecules per unit of volume. Now, if A follows a diffusive movement with a mean free path much lower than R, the above formula cannot be used since the particle will repeatedly bump against the *same* target B until it moves by a distance of order of R. The root mean square displacement during time t is $(6Dt)^{1/2}$. The time required to span a sphere of radius R is thus of order of $R^2/6D$. Since the relative diffusion coefficient of molecules A and B is $D=D_A+D_B$ (the average square of the relative displacement of A and B during a given period of time is the sum of squared displacements, due to random orientation), the expected rate constant of association of A and B should be :

$$d^+ = [(4\pi/3) R^3] / [R^2/6 (D_A+D_B)] = 8\pi R (D_A+D_B) \qquad (26)$$

Note that a factor of 1/2 should be added if molecules A and B are identical, in order to avoid double counting collisions. An important (and well known) consequence of this formula is that the association rate is only weakly dependent on the molecule size. Indeed, if molecules A and B are spheres of identical radius R/2, the diffusion coefficient $D_A$ or $D_B$ is given by $kT/6\pi\mu R$ (as obtained by combining Einstein's and Stokes' formulae ; $\mu$ is the medium viscosity). In aqueous solution, $\mu$ is about 0.001 Pa.s and $d^+$ is equal to $8kT/3\mu = 1.10 \; 10^{-17}$ molecule$^{-1}$·m$^3$·s$^{-1}$ = $6.64 \; 10^9$ M$^{-1}$ s$^{-1}$.

*The standard Smoluchowski theory*. The basis of many recent theoretical studies remains the classic work by Smoluchowski (1917 ; see Bongrand et al., 1982, for a brief summary). This theory was elaborated to account for the rate of coagulation of suspensions of colloidal spheres driven by brownian motion. A central immobile particle is considered as a "perfect sink", and the flux of diffusing particles is calculated by solving the standard diffusion equation :

$$\partial c/\partial t + D \Delta c = 0 \qquad (27)$$

where c(r,t) is the concentration of diffusing particles at time t and distance r from the sink center. Using the simple form of the Laplacian operator $\Delta$ under spherical coordinates for a function depending only on variable r (i.e. (1/r)d{rdc/dr}/dr), equation (12) is readily solved using as boundary conditions c(R,t)=0 and c($\infty$,t)=$c_0$ (i.e. the initial concentration). The total number of collisions affecting the central particle between time 0 and time t is :

$$N = 4\pi D R c_0 t [1 + 4R/(\pi Dt)^{1/2}] \qquad (28)$$

There remains to discuss the relevance of this equation to the rate of encounter between molecules A and B we first considered. There are two important points :

- first, in order that the number of encounters be linearly dependent on time, Dt should be much larger than $16R^2/\pi^2$ (equation 13). The physical meaning of this condition is quite clear : the interaction range R must be substantially higher than the diffusive displacement. It is easily found that this condition is not very restrictive. Suppose A and B are typical proteins (molecular weight 50,000, density 1.3 g/cm$^3$, modeled as spheres of 2.5 nm radius). The relative diffusion constant (D=2 × kT/6π μa, where μ is the medium viscosity) is about $1.8 \times 10^{-10}$ m$^2$/s. The corresponding condition is that t is higher than $6 \times 10^{-8}$s, which is easily satisfied under current experimental conditions. Equation (28) is then equivalent to equation (26).

- Second, there remains to know how long equation (28) may be considered as valid. We may ask what happens when the number of collisions is higher than 1 (the corresponding time is about 10 μs when molecules A and B are spheres of 2.5 nm radius and 10 μM concentration). Two limiting cases may be considered :
    i) if most encounters result in complex formation, equation (28) is no longer valid after the first collision. Thus, as pointed out by Collins and Kimball (1949), the physical significance of the concentration gradient is difficult to understand. However, this difficulty does not invalidate the model, since this relies on the accepted assumption that the "macroscopic" concentration field may be

viewed as a probability density. This point is somewhat clarified by a rigorous discussion presented by Collins and Kimball (1949). See also Berg (1993) for a successful use of a similar kind of reasoning.

ii) if only a small fraction of encounters results in complex formation, our zero concentration boundary condition (c(R,t)=0) is no longer valid. The physical meaning of this situation is that equation (28) gives the number of "first collisions" between molecule A and molecules B diffusing from other parts of the solution. Thus, the "true" collision rate should be equal to the product between N (as given by Equation 28) and the mean number *n* of encounters between A and any molecule B having undergone at least one encounter with A. The problem is that *n* is very difficult to calculate - and even to define - within the framework of the "macroscopic" Smoluchowski theory. Indeed the probability that a molecule B at distance r from A will "encounter" A (i.e. will approach within distance R) is r/R (see. e.g. Berg, 1993). Thus, parameter *n* is highly dependent on the microscopic properties of encounter, since the probability that a particle B leaving particle A will encounter A again may be infinitely high if the jump length (i.e. mean free path) is vanishingly small. Collins and Kimball (1949) suggested that this situation might be managed by replacing the "zero concentration" condition with the following "radiation boundary condition" :

$$c(R) = \gamma \, \partial c/\partial r_{r=R} \tag{29}$$

The justification of this formula is that the particle flux at the boundary is equal to the association rate, which is proportional to the local particle concentration (i.e. c(R)). This condition was shown by Shoup and Szabo (1982) to be consistent with the encounter-complex model, as well as Kramers' reaction rated theory (Kramers, 1940) that will be discussed below. However, we shall describe below another approach that may yield a more intuitive understanding of the process of bond formation, relying on computer simulation. First, we shall describe two further refinements of Smoluchowski theory, assuming high efficiency of bond formation after encounter.

*Interaction potential.* The effect of an interaction between colloid particles was first studied by Fuchs (1934) and a similar reasoning was applied by Debye (1942) to the interaction between charged electrolytes. Assuming the occurrence of a *centrosymmetric* potential U(r) between particles A and B, the relative flux of type B particles around a given particle A is now :

$$\vec{J} = -D \, \vec{\mathrm{grad}} \, c - \frac{cD \, \vec{\mathrm{grad}} U}{kT} \tag{30}$$

the conservation equation is readily solved after replacing c(r) with φ(r) exp(-U/kT). The Smoluchowski expression for the rate constant is replaced with :

$$d^+ = 4\pi DR / R \int_R^\infty \exp(U(r)/kT) dr / r^2 \tag{31}$$

the denominator of eq. 31 is sometimes called the "stability ratio" when applied to colloid suspensions.

*Hydrodynamic interaction.* Another important parameter is the hydrodynamic repulsion between approaching molecules. The friction coefficient for the displacement of a sphere of radius $R_A$ moving towards the center of a sphere of radius $R_B$ with a vanishingly small distance h between surfaces may be approximated as $6\pi\mu\{R_A R_B/(R_A+R_B)\}^2/h$ (Dimitrov, 1983). This phenomenon was accounted for by Spielman (1970) by replacing the constant diffusion coefficient D in equation (30) by a function D(r), yielding for the encounter rate constant :

$$d^+ = 4\pi D_\infty R / R \int_R^\infty D(r) \exp(U(r)/kT / D_\infty r^2 \, dr \tag{32}$$

The problem is that the relatively simple equations (31) and (32) cannot be applied to interactions involving realistic molecules with asymmetrical shape and structure. Also, the association between complementary sites borne by receptor and ligand molecules require that these sites encounter with proper orientation. If alignment is required with angstrom accuracy, corresponding to an angular rotation of order of 1/10 rd for each molecule, viewed as a sphere of 1nm radius, the probability that this condition be satisfied would be of about $(1/400)^2$ (the solid angle subtended by an angle $\theta$ is $2\pi(1-\cos\theta)$). Thus, the expected correction is by no means negligible. In view of the impossibility to obtain analytic formulae for the association rate between realistic molecules, most recent insight was obtained with the powerful method of computer simulation. Before indicating some recent results, we shall briefly discuss the expected properties of surface-bound receptors.

*Expected behavior of membrane-bound molecules.* It may be of interest to discuss the relevance of models concerning soluble molecules to surface-attached receptors. Let us consider two cell membranes bearing adhesion molecules within binding distance. A characteristic diffusion constant D for lateral diffusion is about $10^{-10} cm^2/s$ (as compared to 10,000 fold higher value in fluid phase for a sphere of 2 nm diameter). Thus, the time required for interacting molecules to exhibit a mutual displacement L of the order of molecular length, i.e. 10 nm, is $\approx$ 1 ms (i.e. $L^2/8D$). This is much longer that the typical time for molecular rotations (i.e. about 0.1 µs for domains of flexible protein molecules ; Dandliker and de Saussure, 1970). This suggests that collision efficiency must be substantially higher between flexible membrane-bound molecules than between free molecules, because close contact is maintained for a higher amount of time. Association between flexible bound molecules should thus be essentially considered as diffusion-limited. If molecules are not fully flexible, the acquisition of a bound state is expected to be substantially impaired, and reaction rates should be dependent on the details of molecular shape and dynamics (see Pierres et al., 1998b, for very simplified quantitative estimates of the dependence of binding rate on membrane separation distance).

*Brownian dynamic simulation of ligand-receptor association.* The brownian dynamics algorithm developed by Ermak and McCammon (1978) was used by Northrup et al. (1984) to determine diffusion-controlled association rates between model spheres. The basic principle is somewhat intermediate between the deterministic molecular dynamics simulation (Alder and Wainwright, 1957) and the Monte-Carlo method (Metropolis et al., 1953). Briefly, simulated trajectories of brownian molecules are obtained by starting from a random configuration and subjecting molecules to stepwise displacement that are the sum of i) a deterministic displacement due to electrostatic and hydrodynamic forces and ii) a random displacement generated by a Langevin-type force. The time step may be of order of 1 femtosecond. The forces may be calculated with various degrees of approximation, and more and more realistic simulations are reported due to the increase of computing power, leading to recent studies of interactions between large biomolecules (e.g. the recent report by Gabdouline and Wade, 1997, on the interaction between the enzyme barnase and its ligand barstar). We shall first sketch basic principles, then we shall describe selected informative conclusions.

*Link between simulation and association rate.* The basic idea (Northrup et al., 1984) consists of splitting intermolecular approach in two parts : first, molecules approach at distance r=b such that the interaction potential may be considered as centrosymmetric when r is higher than b, allowing the use of analytical formulae based on Smoluchowski approach. Second, a number of trajectories are simulated with a starting distance b and random initial orientation. The association rate is calculated as the product between the rate of approach at distance b (given by a Smoluchowski constant k(b)) and the probability that a molecule starting from distance b will eventually react rather than departing to infinity. There are some points of interest with this approach :

i) the choice of b is a matter of convenience and does not change the final result. This is easily shown in the case of freely diffusing spheres A and B with 100% reaction probability at distance a. Indeed, it may be shown with simple reasoning that the probability that a diffusing point starting at distance b from a sphere of radius a will encounter the sphere is a/b (Berg and Purcell, 1977 ; Berg, 1993). Thus, using Smoluchowski's formula, the rate constant we expect to obtain with Northrup algorithm is :

$$d^+ = 8\pi Db \times (a/b) = 8\pi Da \tag{33}$$

ii) This approach allows easy account of the possibility that only a fraction of encounters (defined as α) will result in reaction. Northrup et al. (1984) defined the following two quantities : $\beta_\infty$ is the probability that a molecule B starting at distance b from A will encounter A (this is $(R_A+R_B)/b$ in absence of further interaction) and $\Delta_\infty$ is the probability that particles separating after an unsuccessful encounter will recollide rather than escape to infinity. Thus, the probability that a type B molecule will react with A after starting from distance b and undergoing exactly n collisions is simply $\beta_\infty ([1-\alpha]\Delta_\infty)^{n-1}\alpha$. The probability that a reaction will occur whatever the total number of collisions is therefore easily obtained by standard summation of the above geometric series, yielding $\alpha\beta_\infty/(1-[1-\alpha]\Delta_\infty)$. The last step consists of deriving $\beta_\infty$ and $\Delta_\infty$ from the study of finite trajectories. This requires the introduction of a suitable truncating procedure, consisting of interrupting trajectories when the distance is higher than some arbitrary threshold q. We refer to Northrup et al. (1984) for a complete description of this method as well as a successful comparison with analytic formulae.

*Selected results obtained by computer simulation of ligand-protein association.* In their pioneering studies, Northrup at al. (1984) studied the interaction between small spheres (0.05 nm radius). They explored the influence of hydrodynamic and electrostatic forces (1 electronic charge per sphere). First, assuming that spheres were uniformly reactive with 100% reaction efficiency per collision, they obtained quantitative agreement between simulation and analytical formulae. Interestingly, the presence of unit charges of opposite sign on spheres induced sevenfold increase of the association rate. Further, when particle reactivity was restricted on hemispheres, electrostatic attraction nearly abolished the retarding effect of this anisotropy.

In a later study, Northrup and Erickson (1992) explored the influence on the binding rate of a requirement for a proper alignment of interacting molecules. They simulated the interaction between model spheres of 1.8 nm radius bearing four potential contact points distributed as a square of 1.7x1.7 $nm^2$ on a fixed tangential plane. Hydrodynamic and electrostatic forces were neglected. Spheres were considered as bound if the distance between contact points was less than 0.2 nm within each pair. They discriminated between a *collision*, defined as a period of time where the distance between spheres remained comprised between 0.2 and 0.4 nm, and an *encounter*, comprising all collisions occurring between the first collision and the separation at a distance of 20 nm, where the probability of further collision was low. A key finding was that the mean duration of a collision was 0.38 ns, whereas an average encounter involved 9 collisions and lasted 6.3 ns, i.e. more than the correlation time for sphere rotation (which was 5.3 ns). Thus, while the Smoluchowski association rate was estimated at $3.8\times10^8$ $M^{-1}s^{-1}$, the calculated reaction rate was $2\times10^6 M^{-1}s^{-1}$, a much higher value than obtained by multiplying the Smoluchowski constant by the geometric probability that collision occurred with proper alignment.

In an other study, Brune and Kim (1994) tried to model enzyme-substrate interaction by considering the association between a model "cleft enzyme" made of two bound spheres and a cylindrical ligand. They estimated the displacement velocity of a particle of mass M along a given axis as $(kT/M)^{1/2}$, according to Maxwell velocity distribution law. They concluded that hydrodynamic forces generated a torque as high as $1.5\times10^{-19}$ N.m tending to help the cylinder approach the model cleft with proper orientation. This value was more than 100 fold higher than the electrostatic torque calculated by assuming a dipolar moment of 300 debye (about $5\times10^{-27}$ C.m), considered as a relatively high value for actual proteins. The conclusion was that hydrodynamic forces might steer interacting ligand and receptors into a proper configuration, thus accounting for the high association rate of many enzyme-substrate couples without a need for electrostatic guidance. A later simulation was performed on a similar system by Antosiewicz and McCammon (1995). They observed that the mean approach velocity was of order of 11-16 cm/s, i.e. much lower than estimated by Brune and Kim. They concluded that hydrodynamic forces had only a moderate influence on association rates, which might be notably lower than that of electrostatic interactions.

Kozack et al. used computer modeling to study the importance of electrostatic steering in the interaction between a monoclonal antibody and hen egg lysozyme. They used X ray structure data to model proteins and aforementioned OPLS functions to model short range intermolecular forces, as well as non linear Poisson-Boltzmann formalism for long range electrostatic interactions. They obtained semi-quantitative agreement with experimental data since the calculated association rate decreased from $8.6\times10^6$ M$^{-1}$s$^{-1}$ to $8\times\times10^5$ M$^{-1}$s$^{-1}$ when electrostatic steering was inhibited by increasing the ionic strength. Experimental constants were $1\times10^6$ M$^{-1}$s$^{-1}$ and $3\times10^5$ M$^{-1}$s$^{-1}$ respectively.

Recently, Gabdouline and Wade (1997) reported a simulation of the interaction between the enzyme barnase and its substrate barstar. The experimental association rate is particularly high ($5\times10^9$ M$^{-1}$s$^{-1}$) due to electrostatic forces, since the kinetic constant exhibits 20fold decrease when electrostatic forces are screened by increasing the ionic strength from 50 mM to 500 mM. Also, a series of mutants were prepared with a rate constant varying over two orders of magnitude (Schreiber and Fersht, 1996), thus making it an attractive challenge to reproduce the relative change of association rate associated to a known interchange of individual aminoacid residues. Gabdouline and Wade neglected hydrodynamic forces, since these forces were supposed to be similar on all mutants. Proteins were very carefully modeled by using crystallographic data, combined with free energy minimization to determine the precise position of individual atoms. Interaction forces were calculated with accepted values of atomic radii and partial atomic charges (the authors used aforementioned OPLS set of parameters derived by Jorgensen and Tirado-Rives, 1988). The authors were partially able to reproduce the relative differences of association rates between different proteins, but a major point that they rightly emphasized was the high dependence of estimated kinetic constants on the definition of the "encounter complex". Three criteria were tested to define protein "contact" : i) the root mean square distance between the coordinates of individual barstar atoms in a given position (relative to barnase) and the expected values of these coordinates for a bound barstar molecule (as deduced from crystallographic data) must be lower than a threshold value. Reasonable rate constants were obtained for values of this threshold ranging between 0.5 and 0.65 nm. However, the calculated rate constant k was highly dependent on this parameter, since it exhibited 100 fold variation when the threshold distance increased from 0.4 to 0.9 nm. ii) The calculated free energy of the barstar-barnase complex in a given conformation must be lower than some arbitrary number (a threshold of - 12 kT was convenient for the wild-type molecules, not all mutants). iii) The encounter complex might be considered as formed when two couples of atoms of interacting molecules (found to be in contact in the cristallographic complex) were less than 0.625 nm apart. The latter criterion gave optimal results, but discrepancies were found for some mutants.

*In conclusion*, only recently reasonable simulations of interactions between realistic large molecules were reported. Results are consistent with the view that the high association rate of biological ligand and receptor molecules might be due to long range steering, e.g. by electrostatic interactions, as suggested by sensitive experimental studies. It may be hoped that an accurate understanding of these interactions will be achieved in the near future.

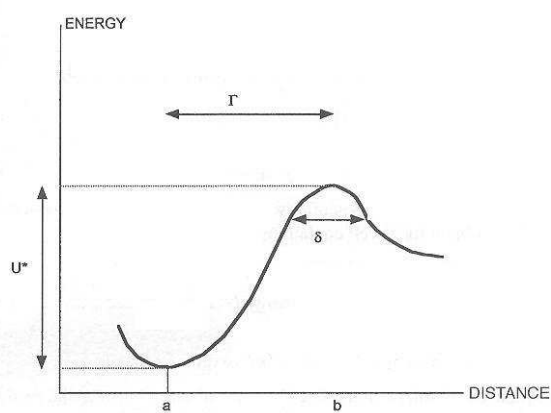

**Figure 8. Simplified model for estimating reaction rates**. It is assumed that the escape of a particle trapped in a potential well (a) follows a unique 1-dimensional path with a major energy barrier corresponding to the transition state (b). The binding range Γ may be defined as the distance between the potential minimum and the maximum corresponding to the transition state. The height U* and width δ of the energy barrier are relevant parameters in addition to the width of the potential well.

*4.2.2 - Dissociation rate*.

It may seem surprising to discuss separately the calculation of association and dissociation constants, since in both cases we are dealing with the problem of diffusion in presence of forces. However, it appeared preferable to sacrifice logics for the sake of clarity.

Early models of particle escape from potential wells were elaborated in the 1930s by Eyring (1935) and Kramers (1940) and much later work was reviewed by Hänggi et al. (1990). The development of experimental methods allowing direct observation and manipulation of the dissociation of invididual bonds generated a broadening of interest of the physical communities in these problems (Bruinsma, 1997) and theoretical studies on the force dependence of dissociation rates (Grubmüller et al., 1996 ; Evans and Ritchie, 1997 ; Izrailev et al., 1997 ; Balsera et al., 1997). We shall first recall simpler models, then we shall describe more recent approaches aimed at providing a link between experimental studies and basic theory.

*Transition state theory.* Eyring's transition state theory (Eyring, 1935) is described in many standard treatises (see e.g. Eyring et al., 1944 ; Hill, 1960, pp 199-200). The first step consists of reducing the model to the escape of a particle (Figure 8) from a potential well following a one-dimensional path. This is justified by normal coordinate analysis and the assumption that there is a *single preferred path* for the reaction. The second step consists of assuming that the reaction rate is essentially limited by the time needed by the particle to reach the "transition state" B : it is thus assumed that when this state is reached, the reaction will proceed with 100 % efficiency (otherwise, an *ad hoc* coefficient must be added). Third, it is assumed that many interactions must occur, leading to thermodynamic equilibrium, before the particle may reach B (this means that the height of the energy barrier $U^*$ must be much higher than $kT$). Using canonical formalism the probability density of finding the system at coordinate x with momentum p is thus proportional to $\exp(-[U(x)+p^2/2m]/kT)$, where m is the effective mass (this should be the reduced mass of the ligand-receptor system if molecules are considered as rigid).

Now, the flux J of particles reaching the transition state is simply the product of the particle concentration near $x_B$ (Figure 8) and the mean velocity of particles moving towards the barrier :

$$J = c(x_B) \int_0^\infty (p/m) \exp(-p^2/2mkT) dp / \int_0^\infty \exp(-p^2/2mkT) dp \qquad (34)$$

and the concentration $c(x_B)$ is given by :

$$c(x_B) = \exp(-U^*/kT) dx / \int \exp(-U(x)/kT) dx \qquad (35)$$

Approximating U as an harmonic potential with force constant $\lambda$ and characteristic pulsation $\omega = (\lambda/m)^{1/2}$, we obtain for a well containing a single particle :

$$J = \omega/2\pi \exp(-U^*/kT) \qquad (36)$$

Now, if we add a distractive force F, the energy $U(x)$ is replaced with $U - Fx$, and the flux becomes :

$$J = \omega/2\pi \exp(-[U^*-F\Gamma]/kT) \exp(-F^2/2\lambda kT) \qquad (37)$$

where we introduced the interaction range $\Gamma$, i.e. the distance between the particle equilibrium position and the position corresponding to the transition state. If $F\Gamma$ is much lower than $U^*$, the last exponential may be discarded (we write $F\Gamma << U^* \sim \lambda\Gamma^2$, and we multiply both sides with $F/\lambda\Gamma$, thus concluding that $F\Gamma$ is much higher than $F^2/\lambda$). Thus, Bell's equation may be considered as an approximation of eq (37).

*Kramers' model.* Kramers (1940) elaborated a model allowing a more precise discussion of the dependence of escape rate on viscosity and temperature. He also considered the 1-dimensional escape of a particle from a potential well as sketched on Figure 8, he wrote the motion equation :

$$m\, d^2x/dt^2 = -f\, dx/dt - dU/dx + X \tag{38}$$

where X is the random Langevin force and f is the friction constant. Kramers then defined as $B_\tau$ the impulsion generated by X during a time interval $\tau$ much higher than the correlation time of X, but lower than the time required for substantial variation of dx/dt. Introducing the density distribution function
$\phi(dx/dt, x, t)$ for B, he obtained the following Fokker-Planck type equation for the density distribution $\rho(dx/dt, x)$ :

$$\partial\rho/\partial t = (dU/dx)\partial\rho/\partial\dot{x} - \dot{x}\,\partial\rho/\partial x + f\partial/\partial\dot{x}(\dot{x}\rho + kT\,\partial\rho/\partial\dot{x})\,d\rho \tag{39}$$

where $\dot{x}$ is dx/dt. Kramers was able to obtain limiting expressions for the escape rate corresponding to low and high viscosity. He concluded that the transition state method should give approximatively correct results within a wide range of f (or viscositiy) values. See Hanggi et al. (1990) for more recent references on the numerical solution of equation (39).

*Simple use of Smoluchowski equation.* We start from the one dimensional form of equation (30). The diffusion current J is :

$$J(x) = -D\, \partial c/\partial x - (Dc/kT)\, \partial U/\partial x \tag{40}$$

under stationnary conditions ($\partial c/\partial d = 0$), J(x) must be independent of x according to the conservation equation (div **J** + $\partial c/\partial t$ = 0). Equation (40) is readily solved, using as boundary conditions c(b) = 0 and $\int_a^b c(x)dx = 1$ :

$$J = D/\int_a^b \exp(-U(x)/kT)\left[\int_x^b \exp(U(y)/kT)dy\right]dx \tag{41}$$

The time for escape is simply 1/J. If we assume that the dominant part of the first integral is contributed by the neighbourhood of x=1, where U(x) is about $\lambda(x-a)^2/2$, and the second integral is mainly contributed by the neigbourhood of the transition state, where U(x)≈U* and the peak width is about $\delta$, we obtain :

$$1/J \approx \delta\, (2\pi kT/\lambda)^{1/2} \exp(-U^*/kT) \tag{42}$$

*Mean first passage time.* An interesting approach to the escape time is the concept of mean first passage time (denoted as W(x)) for a particle located at point x between a and b (see Szabo et al., 1980 ; Hanggi et al., 1990). The basic equation is obtained by considering a particle starting from position x at time 0. A short time $\tau$ later, the distribution of particle concentration is $c(y,\tau)$. Thus, we may write :

$$W(x) = \tau + \int W(y)\, c(y,\tau)\, dy \tag{43}$$

Now, we take the first derivative of equation (43) with respect to $\tau$ and we replace $\partial c(y,\tau)/\partial\tau$ with the divergence of the particle flux, following Smoluchowski equation. We obtain :

$$0 = 1 + D \int W(y)\, [\partial^2 c/\partial y^2 + c/kT\, U'' + \partial c/\partial y\, U'/kT]\, dy \tag{44}$$

were ' and " mean the first and second derivatives. Now, at time zero c(x,t) is simply Dirac "function" $\delta_x$. The integral in (44) may be readily calculated with two integrations by part (we may also notice that $\delta_x$ is a distribution, and use the known rules of the derivation of distributions ; Schwartz, 1966). We obtain :

$$W'' - (1/kT) U' W' = -1/D \qquad (45)$$

(assuming D is constant - a generalization of eq. (45) would be easily obtained). Equation (45) may then be solved with the following boundary conditions : $W(b) = 0$ (which is obvious) and $W'(a) = 0$ (reflecting boundary - see e.g. Hanggi et al., 1990 ; Berg, 1993). The solution is :

$$W(a) = (1/D) \int_a^b \exp(U(y)/kT) \left[ \int_x^b \exp(-U(y)/kT) dy \right] dx \qquad (46)$$

We obtain a result comparable to that yielded by Smoluchowski's approach.

*Models for the mechanical rupture of ligand-receptor bonds.* Experimental results on the rupture of individual ligand-receptor bonds by atomic force microscopy (Moy et al., 1994) were an incentive to extend previous models. Grubmüller et al. (1996) reported a computer simulation of rupture of the streptavidin-biotin bond. They simulated a very stiff cantilever, since the spring constant was 2.8 N/m (i.e. nearly twentyfold higher than in experimental studies). Further, the timescale of force increase was nanoseconds rather than milliseconds. The main results were i) the rupture force increase with increasing loading rate, and the extrapolated value was 280 pN at low rate of force increase. ii) when the force was plotted versus cantilever position, the authors obtained numerous peaks that could be ascribed to individual molecular interactions.

In a later study, Izrailev et al. (1997) simulated the rupture of the avidin-biotin bond during a period of 40- 500 ps, with spring constants ranging between 60 mN/m and 2.8 N/m. They reported rupture forces as high as 450 pN. Also, they presented a theoretical study (based on mean first passage time) to demonstrate that computer simulation limited to nanosecond duration could not reproduce the thermally activated bond rupture requiring milliseconds to occur.

At the same time, Evans and Ritchie (1997) presented a very thorough extension of Kramers's model to force-activated bond rupture. First, they developed the constant flux approach based on Smoluchowski equation and considered the effect of a wide range of disruptive forces on escape from potential wells modeled with power law. Second, they tested their theoretical predictions with a previously published extension of Monte Carlo Method ("smart Monte Carlo simulation", Rossky et al., 1978) that allowed to study the detachment induced by forces ranging over eight orders of magnitude by extending the temporal range of simulations. Also, they presented some comparison of one-dimensional and three dimensional detachment. They were thus able to show that the molecular significance of the rupture force was highly dependent on the timescale and rate of increase of applied force. Thus, viscous drag played a dominant role to retard ultrafast bond rupture by rapidly increasing forces. Finally, they proposed a tentative law for the dependence of avidin-biotin rupture force on loading rate, with an increase from about 100 pN to 400 pN when this rate increased from 1 to $10^{20}$ pN/s. Another interesting result of this study was the demonstration that the time of first passage through the transition state might be much lower than the time of "fairly" definitive escape when the loading rate was low.

### 4.2.3 - *Affinity constant.*

Since the affinity constant is arguably the most important parameter in the characterization of a ligand-receptor interaction, an essential question is to derive this constant from structural data obtained on biomolecules. A full discussion of this problem would not fall into the scope of this review, but it is useful to highlight some conceptual problems that are responsible for some discrepancies found in present literature. We refer the interested reader to a very informative review by Gilson et al. (1997) for more details. Starting from Equation (5), the problem consists of evaluating the standard Gibbs free energy of reaction $\Delta G°$.

First, if A and B are *rigid bodies*, it seems warranted to split $\Delta G°$ into the following components:
- the intrinsic contribution $\Delta G^i$ results from the association between complementary sites with concomitant displacement of solvent molecules. The evaluation of this term was discussed in § 4.1.

- The connection free energy $\Delta G^c$ results from the loss of translational and rotational degrees of freedom resulting from the lumping of two molecules (AB) into a complex (AB). For the sake of simplicity, we shall assume that A is a fixed receptor. An extreme view would be to assume that $\Delta G^c$ is the sum of the translational and rotational free energies of molecule B. In gas phase, the classical partition function associated to the translational motion of a free point particle of mass m in volume V is:

$$Z = (V/h^3) \iiint \exp(-\{p_x^2 p_y^2 p_z^2\}/2mkT)\, dp_x dp_y dp_z \qquad (47)$$

were p stands for the momentum along direction x, y or z and h is Planck's constant. Further, the translation and rotation of a fully asymmetric molecule (Hill, 1960) contribute the following quantity to the chemical potential $\mu = \partial G/\partial n$ (recall that Z is $- kT \ln F$) :

$$\mu = -kT \ln[\{2\pi mkT/h^2\}^{3/2}/v^\circ] - kT - kT \ln[\pi^{1/2} (8\pi^2 I_A kT/h^2)^{1/2} (8\pi^2 I_B kT/h^2)^{1/2} (8\pi^2 I_C kT/h^2)^{1/2}]...$$

$$+ kT \qquad (48)$$

where $I_A$, $I_B$ and $I_C$ are the principal moments of inertia, k and T are Boltzmann's constant and the absolute temperature respectively, $v^\circ$ is the volume per molecule under standard conditions. The first term on the right hand side is the classical Sackur-Tetrode formula for the translational free energy. If we apply this formula to an average protein molecule of molecular weight 50,000 and radius 2.5 nm, the translational free energy is 4.7 kcal/mole per translational degree of freedom and 5.7 kcal/mole per rotational degree of freedom, leading to a total connection free energy of 31.8 kcal/mole, which is much higher than e.g., current values of total standard free energies of ligand-receptor association. Note that the above value is only weakly (i.e. logarithmically) dependent on molecular mass.

Now, as convincingly emphasized by Gilson et al. (1997), the binding of molecule A does not result in complete loss of translation and rotation, only these free motions are replaced with vibration modes. The free energy penalty associated to the loss of translation is thus of order of $kT \ln(v/v^\circ)$, where a convenient order of magnitude for v is the region where the potential energy is less than the minimum value by less than kT. Since the standard volume $v^\circ$ per molecule in a 1 molar solution is 1.6 $nm^3$, the translational free energy increase associated to the confinement of the molecule in a region of space of 0.05 nm dimension (corresponding to the root-mean-square value of atomic deviations in a protein cristal ; e.g. Karplus and McCammon, 1981) would be 5.7 kcal/mole. This order of magnitude is consistent with different reported estimates (Jencks, 1981 ; Vajda et al., 1994). Note that this calculated value is independent of the mass of reacting molecules, in constrast with formula (48).

When interacting molecules are flexible, the reaction free energy may be written as the sum of three components, corresponding to the intrinsic binding energy $\Delta G^i$ (i.e. the direct interaction between atoms of interacting molecules), the internal modifications of reagents $\Delta G^{int}$ (i.e. the modifications of the internal free energies of A and B. This may be important since total conformational energies are much higher than standard free energies of interaction), and the contribution of the loss of overall molecular motions $\Delta G^{ext}$ (corresponding to the external term defined by Gilson et al., 1997 ; as clearly emphasized by these authors, there is some arbitrariness in the separation of internal and external components ; also, the bound state may be difficult to define in loosely bound complexes). In their study made on 13 different complexes between oligopeptides and histocompatibility molecules, Vajda et al. (1994) concluded that flexibility might contribute 30-50 % of the free energy change.

In conclusion, despite some interesting studies on model compounds (see Brady and Sharp, 1997), the derivation of binding free energies from structural data, based on extensive computer simulation, is hampered by the complexity of macromolecules, and the fact that final energies are a sum of much larger components of opposite sign. Clearly, more work is required before the interaction energy between two protein molecules can be deduced from the mere primary sequences of aminoacids.

**CONCLUSION**

At least three main advances were achieved during the last few years : first, as described in section 2, accurate information became available on the structure of several representative ligand-receptor complexes, including the location of individual atoms and solvent molecules in contact areas, as well as detailed comparison of the conformation of free and bound molecules. Second, as reported in section 3, powerful experimental approaches yielded quantitative information of the natural lifetime and force dependence of individual bonds formed between different couples of biomolecules. Third, as summarized in section 4, computer simulations were recently performed to study interactions between realistic macromolecules in aqueous environment, although the duration of the simulation was much shorter than a typical biological process.

Now, there remains to ask what future developements and applications can be expected from this basis.

First, the analysis of important cell functions such as adhesion, spreading or migration, as described in the introduction of this review, should be markedly facilitated by our improved understanding of the behaviour of adhesion receptors. Indeed, the previous development of quantitative models was impaired by our incomplete knowledge of even the order of magnitude of parameters such as bond lifetime, mechanical strength or kinetic of association on a surface.

Second, computer simulation procedures allowing increasingly accurate prediction of the interaction properties of two given biomolecules of known structure may prove helpful e.g. for drug design, in order to obtain inhibitors or competitors of cell receptors. However, some progress is still needed, since many present simulations rely on experimental cristallographic structures of explored molecules. However, it is a reasonable hope that the growing reliability of available algorithms will soon allow safe prediction of the structure of chemically defined molecules. However, an important question is to know whether present tools of molecular mechanics, in addition to their predictive power, yield an accurate picture of actual molecular behavior. Indeed, due to the complexity of computational techniques, it is more and more difficult to rule out the possibility that the agreement between calculated and observed energies might reflect a cancellation of errors obtained by *gradual selection of successful algorithms*. Indeed, it is striking to note that empirical potential functions must be selected according to the predictive power rather than the accuracy of representation of actual forces (see Warshel and Åqvist, 1991). Also, as pointed out by Isralachvili (1991), many functions in addition to Lennard-Jones potential might be used to mimick intermolecular forces.

Third, a permanent task of scientists must be to subject basic principles to more and more stringent tests. Thus, it will certainly be useful to use algorithms that were devised to estimate interaction energies in order to estimate binding forces and bond lifetimes, which should be a more stringent way of checking their validity.

Fourth, although it may seem an hopeless task to develop analytic formulae for predicting the association or dissociation of realistic molecules, there are several reasons for keeping an interest in calculations such as are described in section 4 : i) these may be used to test the reliability of computational methods, by comparing analytical results and numerical estimates when these are applied to simple systems. ii) A clever use of analytical formulae may substantially decrease computational load, thus increasing the range of application of presently available simulation procedures. iii) finally, even approximate analytic formula may be used as a basis to obtain an intuitive understanding of the significance of numerical data.

Thus, although we are still a long way from the *ab initio* prediction of the parameters of intereaction between two protein molecules of given sequence, it may be hoped that the continuous lines of research we describe lead us nearer and nearer to this goal.

**References.**